\documentclass[aip,cha,preprint]{revtex4-1}

\usepackage{amssymb}
\usepackage[english]{babel}    
\usepackage[dvips]{graphics}
\usepackage{subfigure}
\usepackage[pdftex]{graphicx}
\usepackage{multirow}
\usepackage{rotating}
\usepackage{picinpar}
\usepackage{color}
\usepackage{lineno}
\usepackage{listings}

\begin{document}

\title{Fractal Descriptors in the Fourier Domain Applied to Color Texture Analysis}

\author{Jo\~{a}o Batista Florindo}
 	     \email{jbflorindo@ursa.ifsc.usp.br}
\affiliation{Instituto de F\'{i}sica de S\~{a}o Carlos (IFSC) Universidade de S\~{a}o Paulo, Av.  Trabalhador S\~{a}o Carlense, 400,
CEP 13560-970 - S\~{a}o Carlos, S\~{a}o Paulo, Brasil} 

\author{Odemir Martinez Bruno}
              \email{bruno@ifsc.usp.br}
\affiliation{Instituto de F\'{i}sica de S\~{a}o Carlos (IFSC) Universidade de S\~{a}o Paulo, Av.  Trabalhador S\~{a}o Carlense, 400,
CEP 13560-970 - S\~{a}o Carlos, S\~{a}o Paulo, Brasil,phone/fax: +55 16 3373 8728 / +55 16 3373 9879}           
                                      
\date{\today}

\begin{abstract} 
The present work proposes the development of a novel method to provide descriptors for colored texture images. The method consists in two steps. In the first, we apply a linear transform in the color space of the image aiming at highlighting spatial structuring relations among the color of pixels. In a second moment, we apply a multiscale approach to the calculus of fractal dimension based on Fourier transform. From this multiscale operation, we extract the descriptors used to discriminate the texture represented in digital images. The accuracy of the method is verified in the classification of two color texture datasets, by comparing the performance of the proposed technique to other classical and state-of-the-art methods for color texture analysis. The results showed an advantage of almost 3\% of the proposed technique over the second best approach.
\end{abstract}


\maketitle


\section{Introduction}

Texture is a visual attribute which is present in most of the nature images. The texture quantification, identification and classification is a very important problem that is defined in computer vision as the study of pixel patterns in an image region. Although this attribute is naturally processed by natural vision and easily comprehended by humans, there is no formal definition for it. Indeed, textures are complex visual patterns formed by arrangements of pixels, regions or even set of patterns formed by other visual attributes, such as shape or color. These patterns can be composed by completely distinct factors, such as pixel organization or even its disorganization. In fact, depending of the context, even the noise can be considered as a sort of texture. These characteristics of the texture attribute make it special and hard to be well defined. 

Along the last years, a lot of methods has been developed for the analysis of textures. Such interest in texture analysis methods may be comprehended by the richness of the texture attribute in images analyzed in pattern recognition problems. 

Basically, the texture analysis methods can be divided into 4 categories \cite{MS98}, that is, the structural methods, in which the texture is described as a set of primitives well defined; statistical methods, in which the texture is represented through non-deterministic measures of distribution; spectral methods, based on the analysis in the frequency domain and model-based methods, based on the mathematical and physical modelling of the texture image.

Among the model-based methods, the fractal model has presented a large projection recently in the description of textures in a wide number of problems \cite{QMACG08,TWZ07,MG05}. Most of these methods employs the fractal dimension direct or indirectly for the representation of the texture. In recent years, however, a family of methods \cite{BCB09,PPFVOB05,BPFC08,FCB10} was developed through extracting a set of features from the fractal modelling, unlike the fractal dimension which is only a unique number. Generically speaking, these methods provide the called fractal descriptors, capable of representing a texture with a higher degree of richness than the simple fractal dimension. 

The ability of fractal features in the description of textures is related to the nature of fractality concept. The fractal dimension measures the complexity of a structure, which in its turn corresponds to physical important properties of a material, like the roughness, the reflectance, among others. Finally, such properties are strong stimuli in our visual identification of texture images, allowing the classfication of objects based on their texture aspects. In this way, fractal theory becomes a worthy tool in the automation of this process.

Actually, the use of term ``fractal'' in such kind of application may bring some controversy, once the analyzed images are not real fractals. Even because fractals are only mathematical entities lacking any perfect representation in the real world. The interested reader may appreciate this discussion for example in a letter exchange involving Avnir et al. and Mandelbrot \cite{ABLM98,M98}. Inasmuch as this debate is not in any way finished, the most acceptable approach in the literature is that presented in Carlin \cite{C00}. There, objects from real world are measured through fractal metrics even when they get away from fractal concept. In this case, the fractal measures acts as a complexity metric of the real world object.

This work proposes a novel technique for the extraction of fractal descriptors based on the Fourier fractal dimension method \cite{R94} from colored textures. The color is an important attribute in texture images, mainly those extracted from natural scenes \cite{MOVY01}. The method consists in representing the color image in a new space color through the linear transform described in Geusebroek et al. \cite{GRAA00}. This transform aims at emphasizing the relation between the pixels color and their spatial distribution. Posteriorly, we apply the method for the calculus of fractal dimension by the Fourier transform. In this method, the dimension is obtained from a curve relating the power spectrum of the Fourier transform and the frequency. Instead of simply using the dimension value, this work proposes the use of the whole curve to provide the descriptors of the texture.

The proposed method takes some important advantages over classical fractal signature techniques like that based on wavelets \cite{MGASR05} or multifractal \cite{H01}. One of such advantages is that the technique here presented gathers information from frequency domain inherently, allowing the capturing of details and patterns which escapes from the conventional spatial analysis. Besides, the space used allows the expression of colors and spatial distribution of pixels as being a related entity. This relation is important in many applications involving color texture. Moreover and not less relevant, the method shows a simple computational implementation and presents a low computational effort.

The performance of the proposed technique is tested in a comparison with other classical and state-of-the-art methods for color texture analysis, namely, chromaticity moments, histogram ratio and multispectral Gabor. The comparison is achieved in the classification of two color texture datasets.

This work is divided into 8 sections. The following addresses the definitions and theoretical aspects of fractal theory. The third section describes the Fourier fractal dimension. The fourth presents the concept of fractal descriptors. The fifth describes the proposed method. The following shows the experiments employed. The seventh section discusses the results and the last one does the conclusions.

\section{Fractal}

The existence of strange objects which do not obey the rules of the traditional euclidian geometry is known from mathematicians from some centuries ago. Nevertheless, the formalization and denomination of such objects is due to Benoit Mandebrot in the 1970 decade \cite{M75}.

A fractal is defined as a set whose Hausdorff-Besicovitch dimension exceeds strictly the topological (Euclidian) dimension. Such fact has as a consequence the fact that the fractals are dynamical systems with infinite complexity. Besides, fractals are self-similar, that is, each part of the object is a similar copy of the whole. It is noticeable that this repetition of patterns along different observation scales is also present in objects found in the nature, like the embranchment of a river, a tree or of the lung alveoli, or still in the nervures of a plant leaf, in a cloud, in a coastline and in many other cases \cite{M75}.

In computational vision problems, we are interested in finding descriptors which characterize the objects in analysis. From the similarities between the aspect of natural objects and the objects studied in fractal theory, researchers started to study the application of a fractal descriptor to objects from the real world \cite{C00}. The most relevant and safe descriptor for this purpose is the fractal dimension.

\subsection{Fractal Dimension}

The literature presents several definitions of dimensions which are generally named as fractal dimension \cite{F86}. Among these, we can cite the Hausdorff-Besicovitch dimension, the packing dimension, the Renyi dimension, the box-counting dimension, etc.

A common point among all these methods is that they are based on the idea of measuring at $\delta$ scale. The analyzed object is measured for different values of $\delta$ and at each different value, the details smaller than $\delta$ are neglected. The fractal dimension thus must express the behavior of the measure as $\delta \rightarrow 0$. In a fractal object, a measure $M_{\delta}(F)$ of a set $F$ must generally obey a power law:
\begin{equation}
	M_{\delta}(F) \sim c\delta^{-s},
\end{equation}
where $c$ is a constant and $s$ is the fractal dimension of $F$. The value of $s$ can therefore be obtained from:
\begin{equation}
	s = -\lim_{\delta \rightarrow 0}\frac{log(M_{\delta}(F))}{log(\delta)}.
\end{equation}
In a general way, $M_{\delta}(F)$ must be a homogeneous function with degree $d$ yielding to the power law
\begin{equation}
	M_{\delta}(F) \sim c\delta^{d-s},
\end{equation}

Here, we describe briefly the development of the first and perhaps most important measure of the fractal dimension, e.g., the Hausdorff dimension. For a general set $F \in \Re^{n}$, the Hausdorff dimension is defined by the following expression:
\begin{equation}
	dim_{H}(F) = \{s\} | \inf \left\{ s:H^{s}(F)=0 \right\} = \sup \left\{ H^{s}(F)=\infty \right\},
\end{equation}
where $H^{s}(F)$ is the $s$-dimensional measure of $F$, defined through:
\begin{equation}
	H^{s}(F) = \lim_{\epsilon \rightarrow 0}{H_{\epsilon}^{s}(F)},
\end{equation}
where
\begin{equation}
	H_{\epsilon}^{s}(F) = inf\left\{ \sum_{i=1}^{\infty}{|U_{i}|^{s}:{U_{i} \mbox{ is an $\epsilon$-cover of F}}} \right\}.
\end{equation}

\section{Fourier Fractal Dimension}
\label{sec:fourier}

The literature still shows alternative definitions for the fractal dimension. Among these definitions, one of the most important is the Fourier fractal dimension \cite{F86}. For the calculus of this dimension we must initially define the Fourier transform of a mass distribution $\mu \in \Re^{n}$, that is, a measure on a bounded subset of $\Re^{n}$ such that $\mu(\Re^{n})$ is positive and finite. The transforms are defined by:
\begin{equation}
	\mathfrak{T}(\mu(u)) = \int_{\Re^n}{e^{ix \bullet u}d\mu(x)},
\end{equation}
\textcolor{black}{where $u$ is a generic subset of $\Re^n$ and $x$ is the Fourier space counterpart of $u$.}

In the following we employ an analogy from the classical mechanics for the definition of the fractal dimension. We use the concept of s-potential of a mass distribution $\mu$ \textcolor{black}{over a point $x$ in $\Re^n$}, given by:
\begin{equation}
	\mathfrak{p}_{s}(x) = \int{\frac{1}{|x-y|^{s}}d\mu(y)},
\end{equation}
\textcolor{black}{where $y$ is an auxiliary variable.}

By still extending the Physics analogy, the potential energy $\mathfrak{e}_{s}$ may be obtained through:
\begin{equation}
	\mathfrak{e}_{s}(\mu) = (2\pi)^{n}c\int{\mathfrak{T}(\mathfrak{p}_{s})(u)\overline{\mathfrak{T}(\mu(u))}du},
\end{equation}
in which \textcolor{black}{$c$ is a constant dependent on $s$ and $n$} and $\overline{x}$ is the complex conjugate of $x$. \textcolor{black}{In this way:}
\begin{equation}\label{eq:energy}
	\mathfrak{e}_{s}(\mu) = (2\pi)^{n}c\int{|u|^{s-n}|\mathfrak{T}(\mu(u))|^{2}du}.
\end{equation}

From a theorem developed in Falconer \cite{F86}, if there is a mass distribution $\mu(u)$ on the set $S \in \Re^n$ for which the expression \ref{eq:energy} is finite for some value(s) of $s$, so the Hausdorff dimension of $S$ has its lower limit in $s$. Particularly, if $|\mathfrak{T}(\mu(u))| \leq b|u|^{-t/2}$, for a constant value $b$, then $\mathfrak{e}_{s}(\mu)$ always converges if $s < t$. The greatest $t$ for which there is a mass distribution $\mu$ on $S$ is called the Fourier fractal dimension of $S$.

For practical purposes, the method for the calculus of the Fourier fractal dimension described in Russ \cite{R94} is applied to a gray-scale (intensity) image $I$. 

\textcolor{black}{In this case, we have a 2D real-valued image $I(i,j)$ with size $N \times N$ and the Fourier transform $\tilde{I}$ is expressed through:}
\begin{equation}
	\tilde{I}(u,v) = \sum_{i=0}^{N-1}{\sum_{j=0}^{N-1}I(i,j)\exp^{-\frac{v2\pi}{N}(ui+vj)}},
\end{equation}
\textcolor{black}{where $u$ and $v$ are respectively the horizontal and vertical frequency. The total frequency $f$ is given by $f = \sqrt{u^2+v^2}$. Another important measure is the power spectrum $P$, given through $P = \tilde{I}^2$}. 

\begin{figure}[!htpb]
	\centering
		\includegraphics[width=0.9\textwidth]{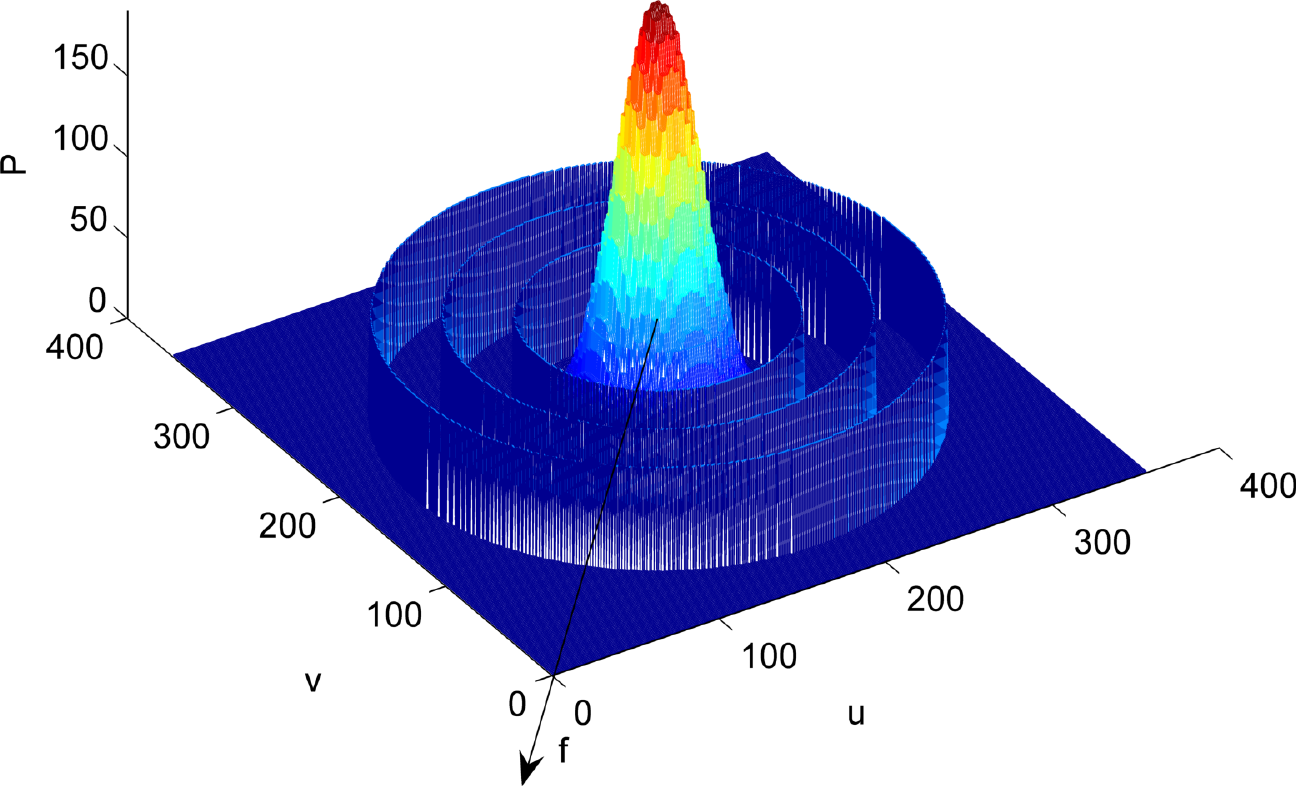}
	\caption{Fourier spectrum is divided into radial rings and the power spectrum $P$ is averaged over each ring. The frequency $f$ corresponds to the average distance of each ring to the center of the spectrum.}
	\label{fig:rings}
\end{figure}

\begin{figure}[htbp]
	\centering
	\begin{tabular}{ccc}
		\includegraphics[width=0.2\textwidth]{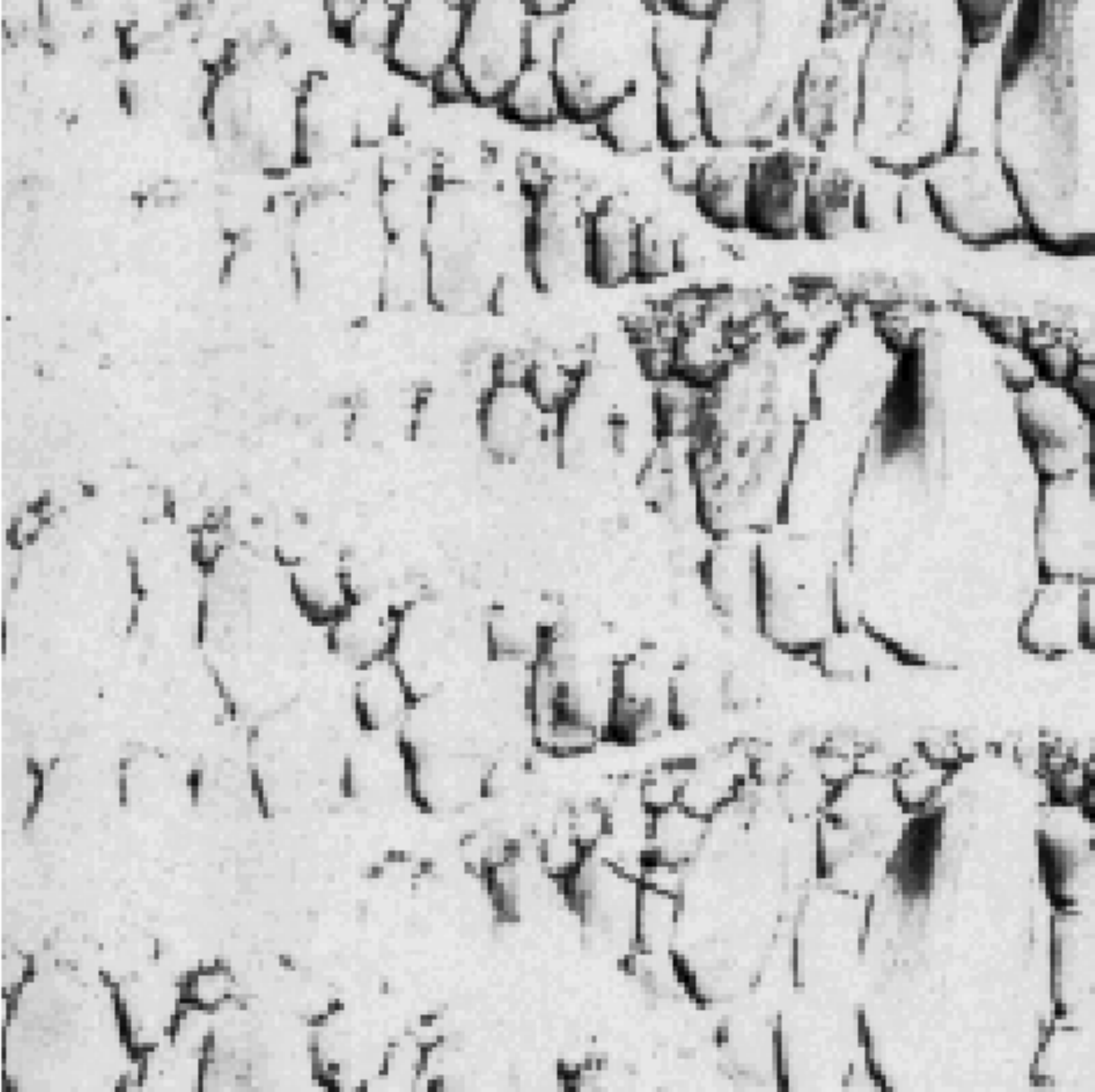} & \includegraphics[width=0.33\textwidth]{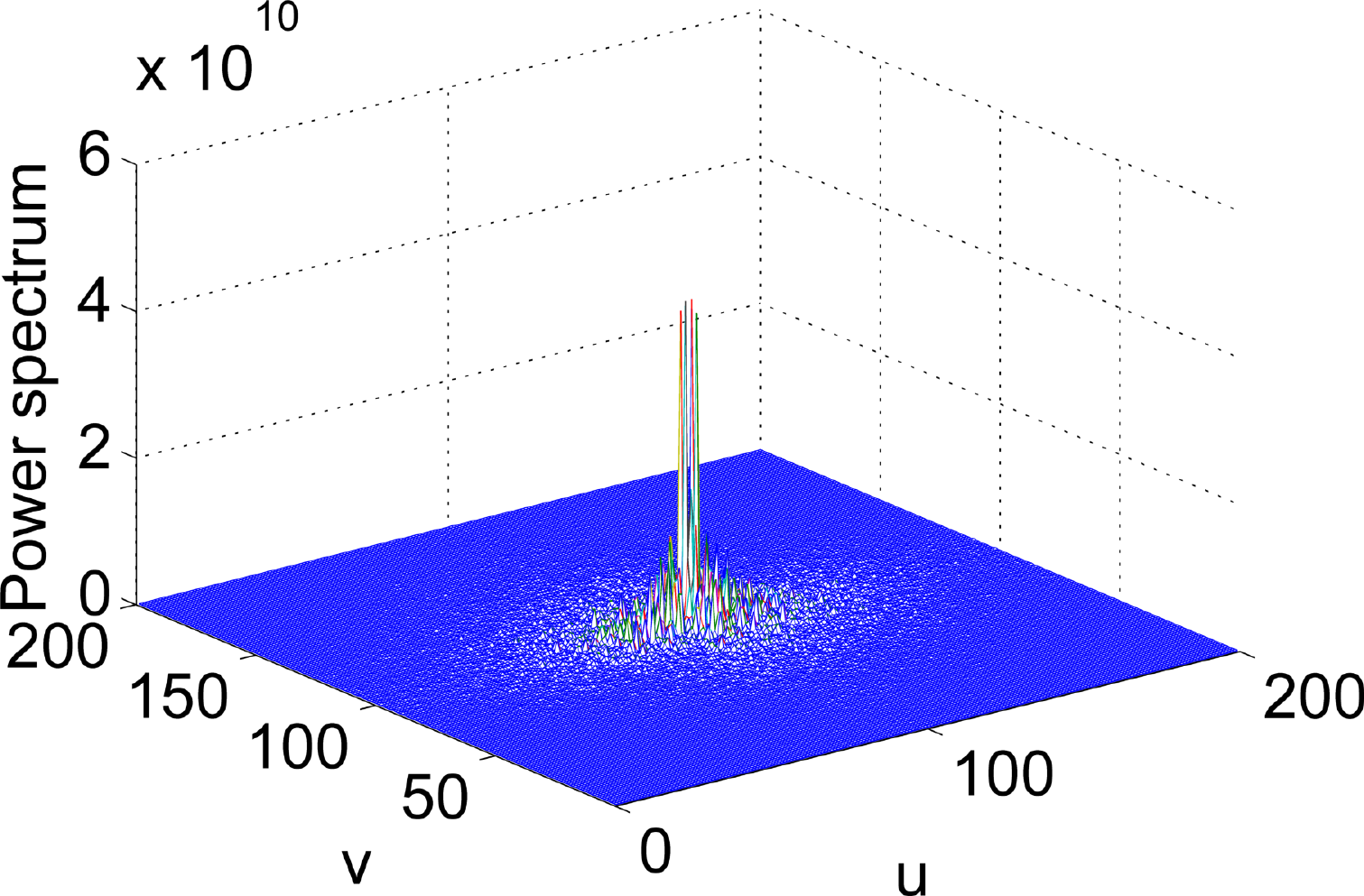} & \includegraphics[width=0.33\textwidth]{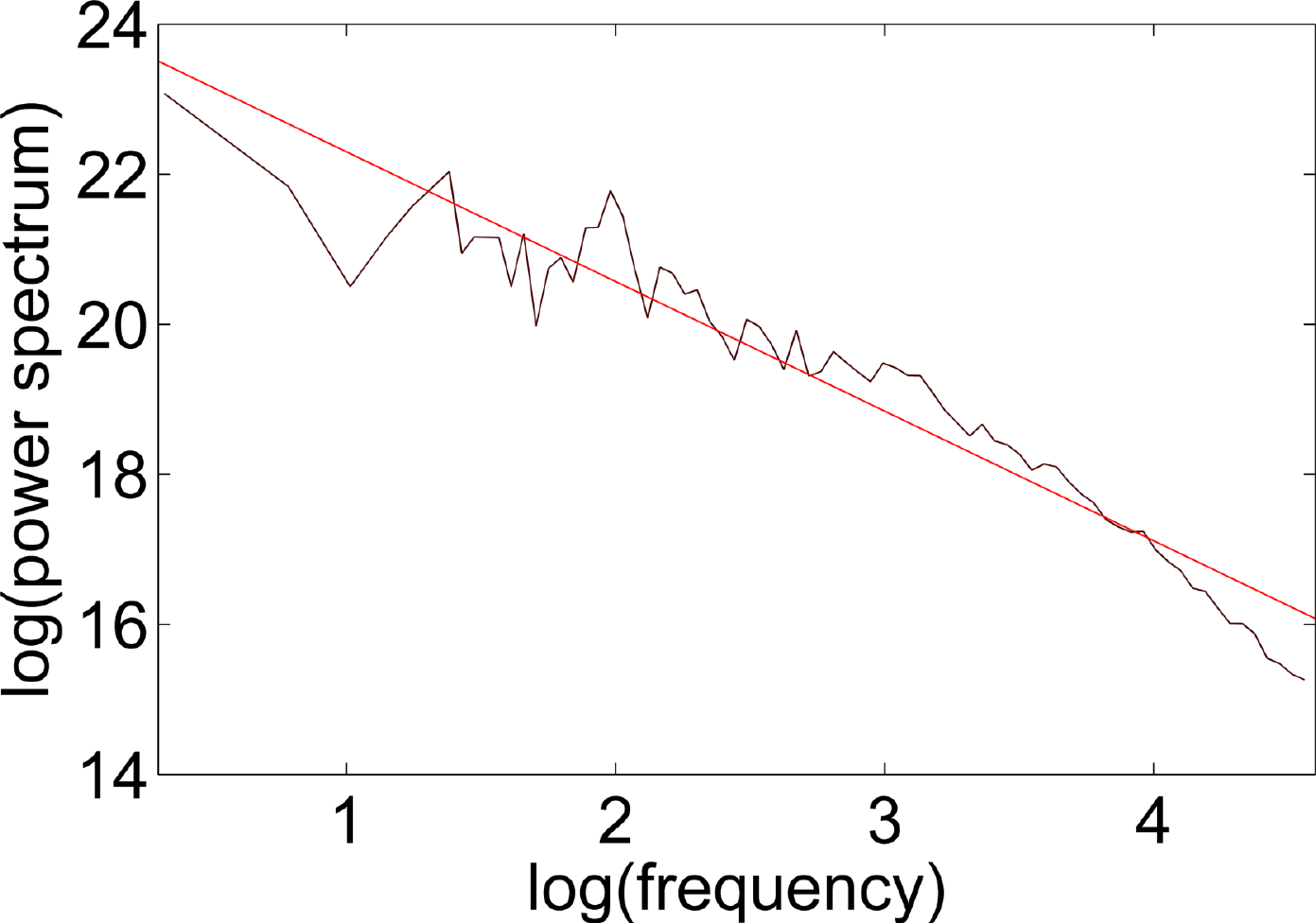}\\
		(a) & (b) & (c)\\
	\end{tabular}
	\caption{Calculus of the fractal dimension of textures. (a) Original texture. (b) Power spectrum. (c) Log-log curver of power spectrum $\times$ frequency.}
	\label{fig:fft}
\end{figure}

Russ \cite{R94} demonstrates that there is an exponential relation between the frequency $f$ and the power spectrum $P$ in the Fourier spectrum of $I$:
\begin{equation}
	P \propto f^{-\alpha},
\end{equation}
He still affirms that the exponentiation parameter $\alpha$ may be used for the estimation of the fractal dimension $D$ of the texture in $I$. In the practice, $\alpha$ is calculated as being the slope of the curve $log(P) \times log(f)$. The dimension is easily estimated through:
\begin{equation}\label{eq:fdruss}
	D = \frac{\alpha+6}{2}.
\end{equation}
\textcolor{black}{In digital images applications, the Fourier transform is calculated by classical optimized techniques, like Fast Fourier Trasnform \cite{B74}. The Fourier spectrum is divided into radial rings (corresponding to frequency bands). Thus, the variable $P$ in the previous expression corresponds to the power spectrum averaged over each ring and the frequency $f$ corresponds to the average distance of each ring from the center of the spectrum. The Figure \ref{fig:rings} illustrates the process.}
The Figure \ref{fig:fft} illustrates the dimension calculated by this practical method.

\section{Fractal Descriptors}

\begin{figure}[htbp]
	\centering
	\begin{tabular}{cc}
		\begin{tabular}{c}
			\includegraphics[width=0.2\textwidth]{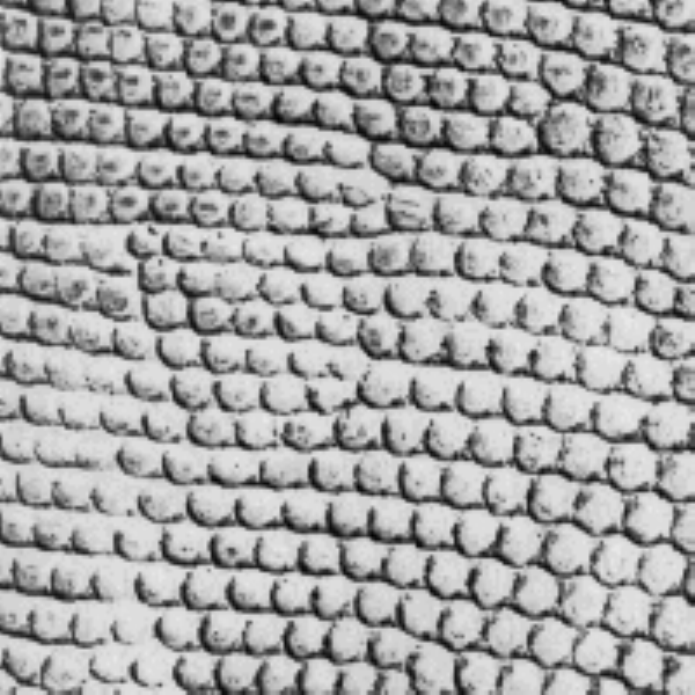}\\
			FD = 1.6291\\	
			\\
			\\	
			\includegraphics[width=0.2\textwidth]{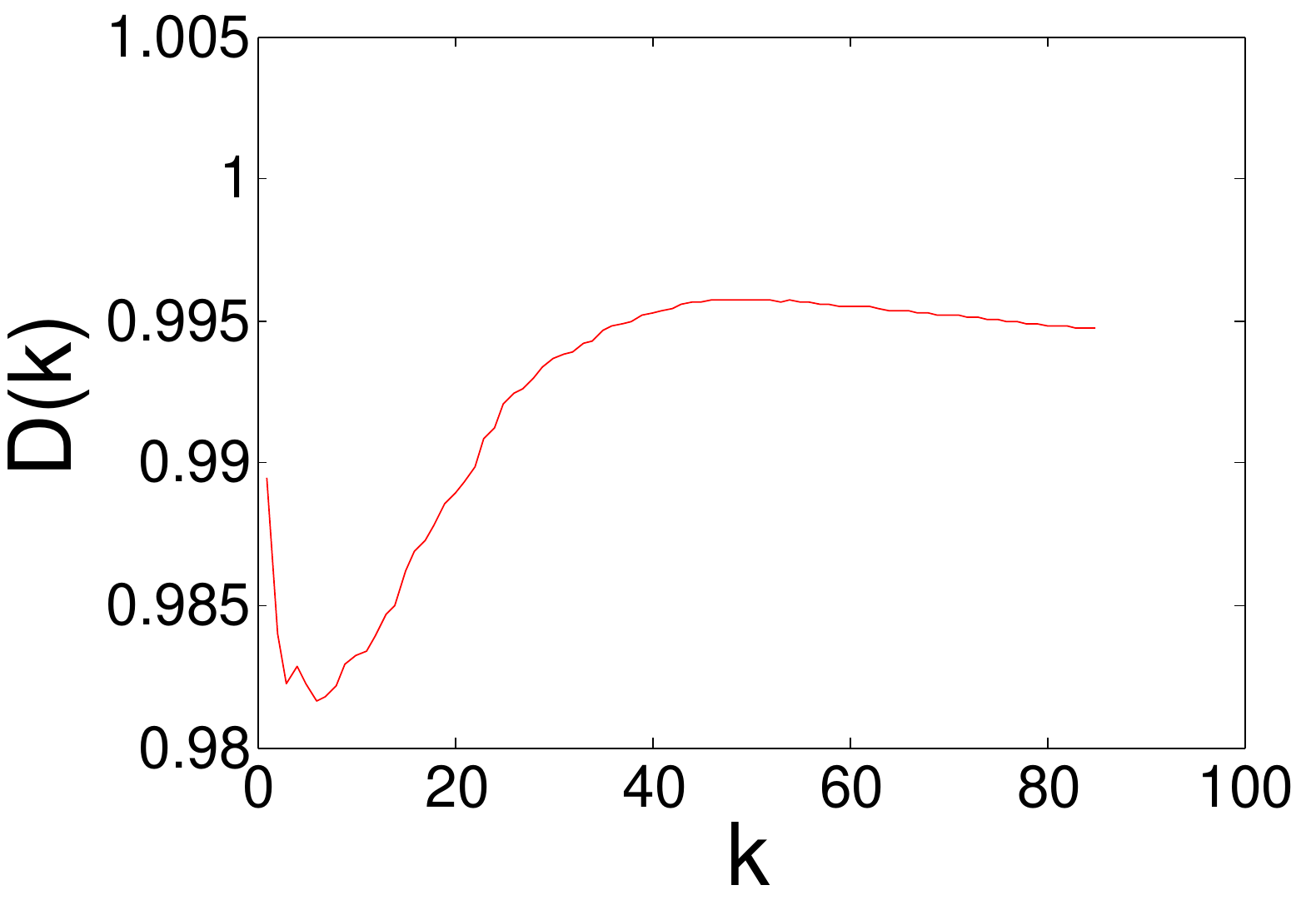}\\ 
			FD = 1.6113\\
		\end{tabular} &
		\begin{tabular}{c}
			\includegraphics[width=0.5\textwidth]{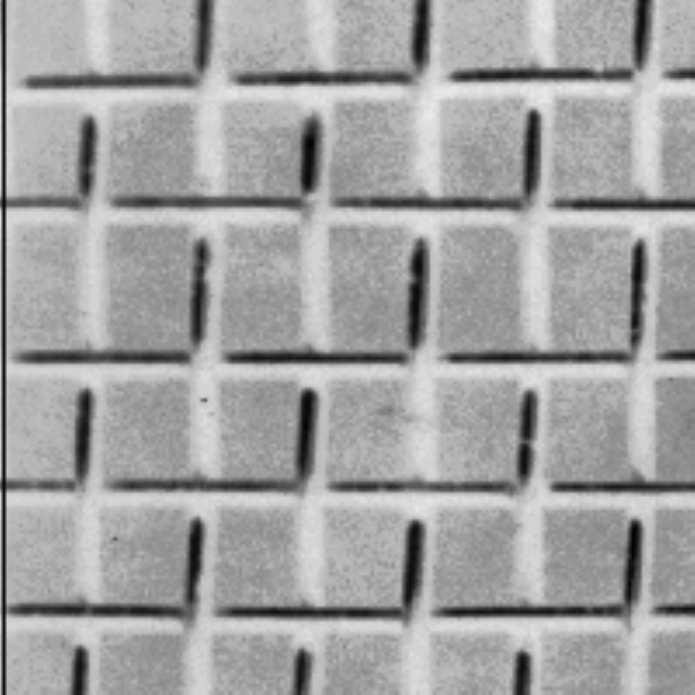}\\
			\includegraphics[width=0.5\textwidth]{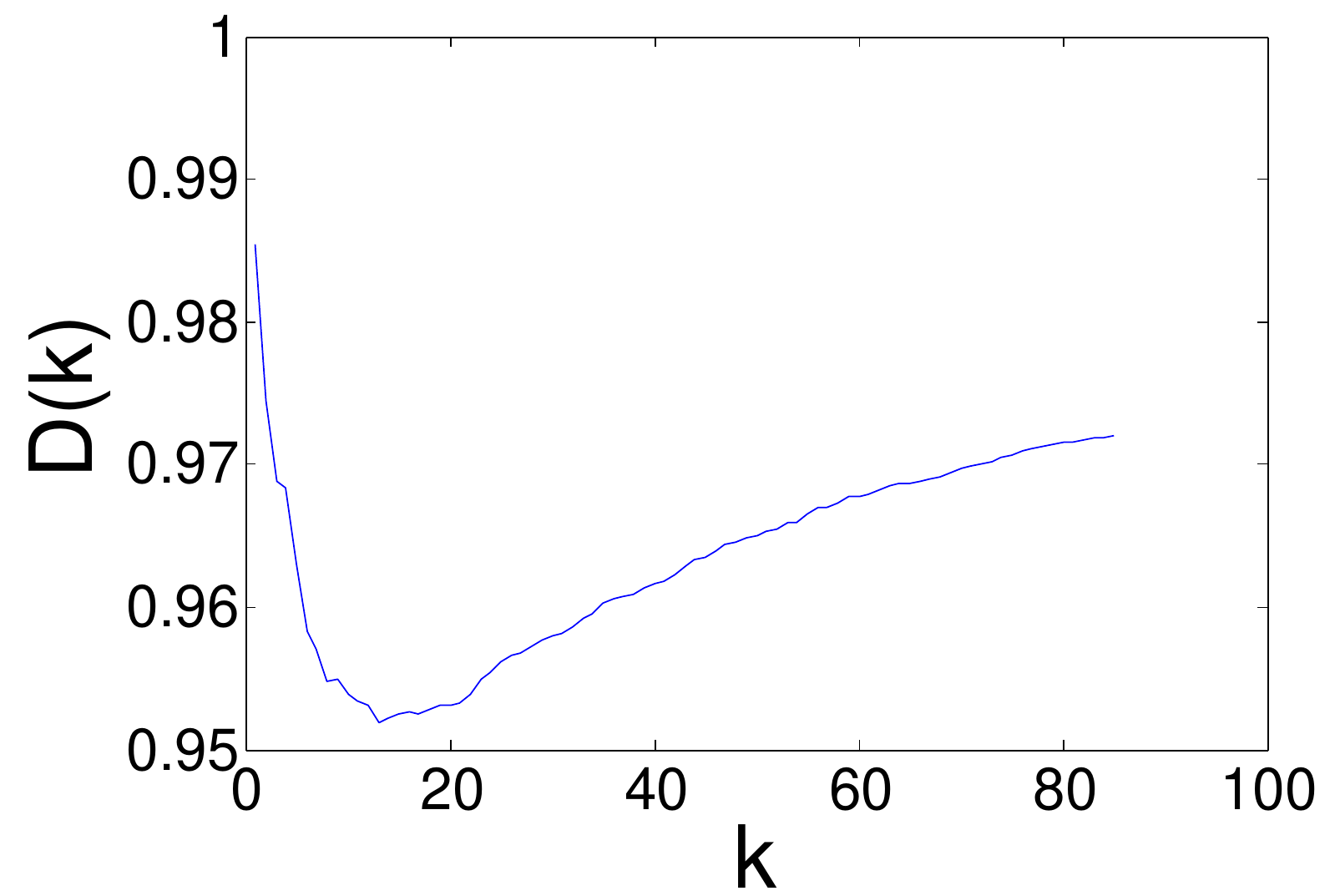}\\ 
		\end{tabular}			
	\end{tabular}	
	\caption{Illustration of the richness of texture fractal descriptors. At left, two textures with similar fractal dimensions. At right, the fractal descriptors for each texture and the clear visual distinction between them.}
	\label{fig:desc}
\end{figure}

Although the fractal dimension is a good descriptor for textures, shapes, contours, etc., they become inefficient in tasks which require a greater precision in the description of the object. Fractal descriptors have arrisen with the aim of filling this gap and provide a more precise technique for the characterization of the image. The Figure \ref{fig:desc} shows graphically the importance of fractal descriptors.

The first known work applying the concept of fractal descriptor is Manoel et al. \cite{MCSM02} which uses the technique named Multiscale Fractal Dimension (MFD) to obtain the descriptors. In this approach, the fractal dimension from the object is inferred at different observation scales. In Manoel et al. \cite{MCSM02}, the authors used the Minkowski sausage method for the calculus of the fractal dimension. In this method, the object is dilated by a variable radius $r$ and the area of the dilated object (number of pixels) is called the dilation area $A(r)$. The fractal dimension is estimated from the slope of the curve $log(A(r))) \times log(r)$. Instead of simply obtaining the dimension, the authors used the whole $log(A(r))$ curve to compose the descriptors by measures extracted from the curve, like the peaks, the area under the curve, etc.

In its turn, in Plotze at al. \cite{PPFVOB05} and Bruno et al. \cite{BPFC08} we have an application of MFD in which the derivative of $log(A(r))$ is used in order to provide the fractal descriptors. Bruno et al. \cite{BCB09} still applies the MFD to the analysis of textures, mapped onto surfaces and using the volumetric Minkowski sausage method.

\section{Proposed Method}

This work proposes a novel method for the extraction of fractal descriptors from colored texture images. 

A lot of methods has been described in the literature for the extraction of features from colored textures with the aim of solving problems like classification and segmentation in different application fields \cite{CD93,P98,JH98}. However, many of such works do not take into account the spatiality of the color, that is, the relation between the color of a pixel and its position in the image or in a specific neghborhood.

In order to deal with this situation, Geusebroek et al. \cite{GRAA00} proposed an interesting method based on physical characteristics of colors. Roughly speaking, the method consists in a linear transform from the original color space into another physical space. In the special case in which the original space is RGB (Red-Green-Blue) like it is our case, the transformation is represented by the simple expression:
\begin{equation}
	\left( \begin{array}{c} \tilde{E}_{\lambda} \\ \tilde{E}_{\lambda\lambda} \\ \tilde{E}_{\lambda\lambda\lambda} \end{array} \right) = \left( \begin{array}{ccc} 0.06 & 0.63 & 0.31 \\ 0.19 & 0.18 & -0.37 \\ 0.22 & -0.44 & 0.06 \end{array} \right)
\left( \begin{array}{c} R \\ G \\ B \end{array} \right),	
\end{equation}
where $R$, $G$ and $B$ are the original color channels and $\tilde{E}_{\lambda}$, $\tilde{E}_{\lambda\lambda}$ and $\tilde{E}_{\lambda\lambda\lambda}$ are the transformed channels. $\tilde{E}_{\lambda}$ corresponds to the convolution of color energy (wavelength) with a gaussian. $\tilde{E}_{\lambda\lambda}$ represents the same convolution with the first derivative of a gaussian, while $\tilde{E}_{\lambda\lambda\lambda}$ is the same with the second derivative of gaussian filter. In Hoang et al. \cite{HGS05}, the authors apply the classical Gabor filters to the transformed color space obtaining interesting results.

In this work, initially we apply the transform described to the texture image. In the following, for each one of the channels $\tilde{E}_{\lambda}$, $\tilde{E}_{\lambda\lambda}$ and $\tilde{E}_{\lambda\lambda\lambda}$ we extract fractal descriptors based on the Fourier fractal dimension described in the Section \ref{sec:fourier}. For the calculus of these descriptors, instead of simply calculate the fractal dimension by the Equation \ref{eq:fdruss}, we use all the values of $\log(P)$ in the curve, that is, the logarithm of the whole power spectrum. Thus, we appy a multiscale transform to the curve, with the aim of capturing the fractal behavior at different observation scales, in a similar manner to that described in Plotze et al.\cite{PPFVOB05}.

Essentially, a multiscale transform is a mapping from the original signal $u(t)$ onto a function $U(b,a)$, where $b$ is related to the original variable $t$ and $a$ is the scale parameter. The literature presents several approaches for the calculus of multiscale transform \cite{W84,CC00}. Based on empirical results we opted for the use of space-scale approach. In this solution, the transform is formally represented through:
\begin{equation}\label{eq:multiscale}
	\{(b,a)|a,b \in \Re, a > 0, b \in \{U^1(t,a)\}_{zc}\},
\end{equation}
in which $._{zc}$ represents the zero-crossings of $.$ and $U^1(t,a)$ expresses the convolution of $u(t)$ with the first derivative of the gaussian $g^1_a$, given by:
\begin{equation}\label{eq:gaussian}
U^1(t,a) = u(t) * g_a^1(t),
\end{equation}
where $a$ denotes the smoothing parameter of gaussian, referenced in the most of textbooks as $\sigma$. The best results were achieved by calculating the derivative through the Fourier property and projecting $U(b,a)$ onto the axis corresponding to $a = 0$. This multiscale representation based on Equations \ref{eq:multiscale} and \ref{eq:gaussian} uses the gaussian kernel concept. It states, using a result from partial derivative theory, that the gaussian filter associated to the first derivative is the unique operation capable of representing an image or signal under different scales without adding any spurious element in the process. More details may be seen in Witkin \cite{W84}

\begin{figure}[!htpb]
	\centering
		\includegraphics[width=0.9\textwidth]{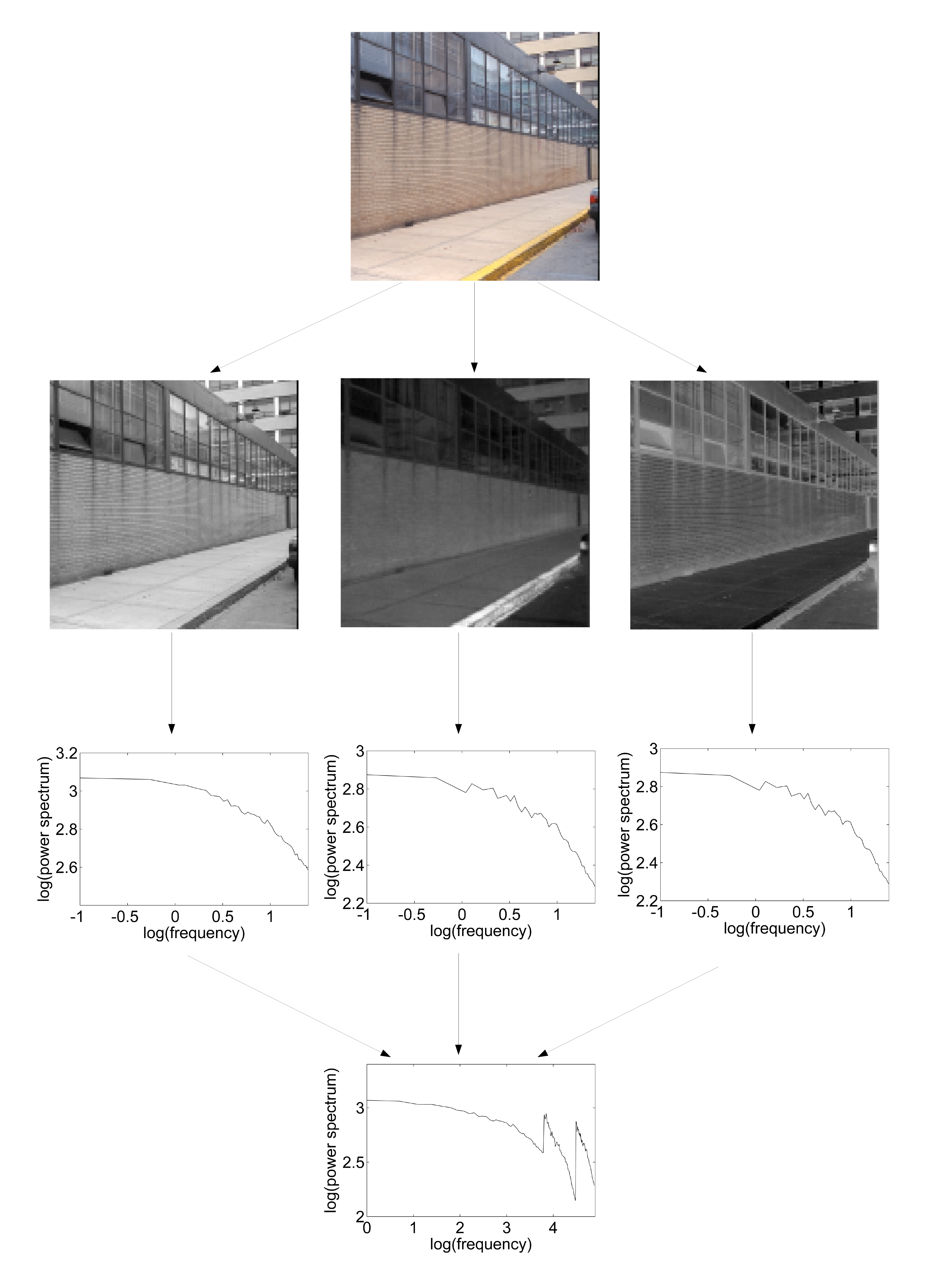}
	\caption{A scheme of the proposed method. From up to down, the original texture, the transformed channels, the Fourier curve for each channel and the final descriptor.}
	\label{fig:metodo}
\end{figure}

Finally, we concatenate the fractal descriptors from each channel, generating the final color Fourier fractal descriptor. The steps of the method are depicted in the Figure \ref{fig:metodo}, while the Figure \ref{fig:algorithm} summarizes the algorithm process.

\begin{figure}[!htpb]
\begin{lstlisting}
Load image
Extract E_lambda channells using the transform matrix
For each channell
	Calculate Fourier tranform
	Divide spectrum into radial rings (frequency band)
	For each ring
		power spectrum = squared magnitude
	end
	Fractal descriptors = multiscale(log(power spectrum)xlog(frequency))
end
Final descriptors = concatenate(fractal descriptors)
\end{lstlisting}
\caption[Título sumário]{Proposed method generic algorithm.}
\label{fig:algorithm}
\end{figure}

\begin{figure}[htbp]
	\centering
		\includegraphics[width=\textwidth]{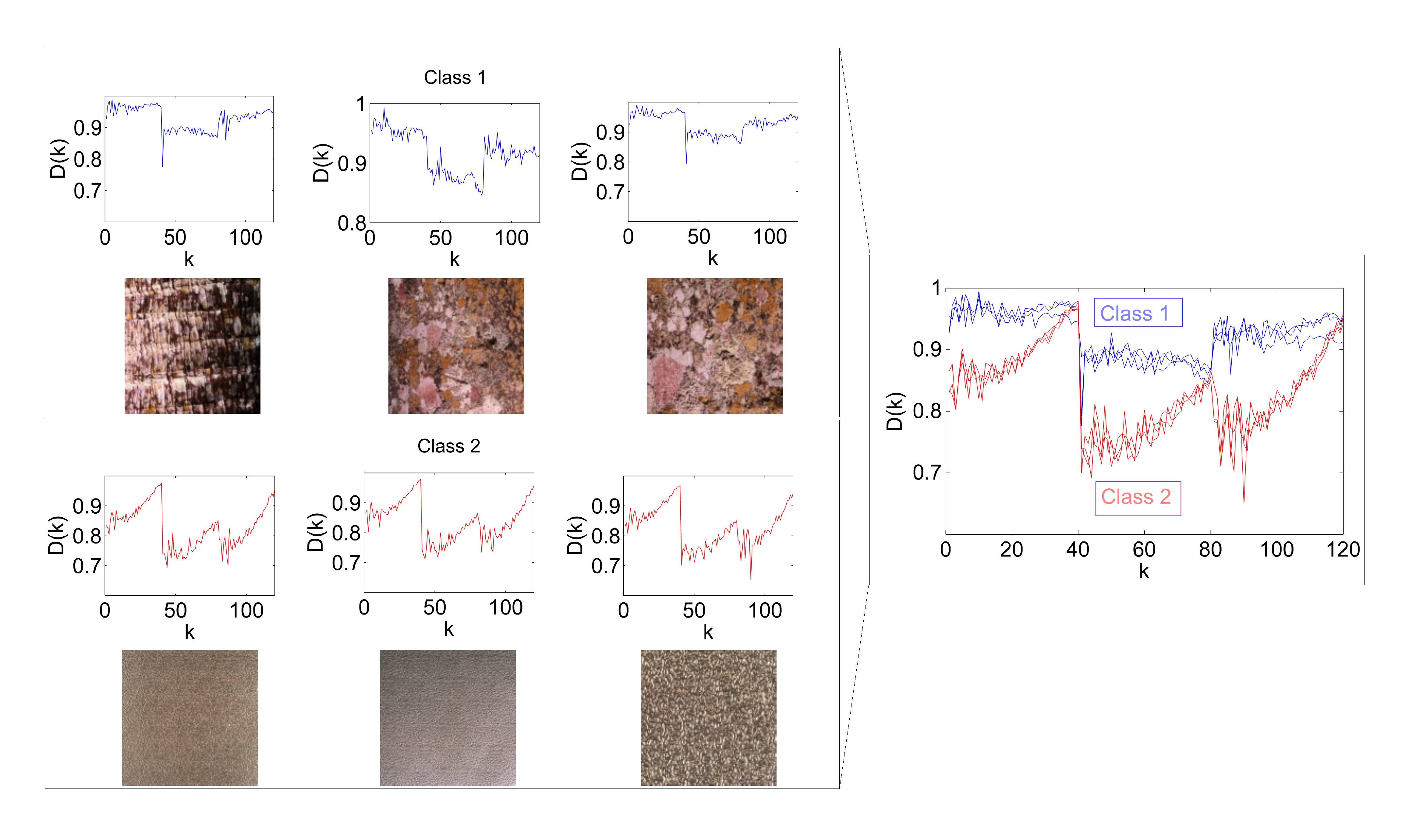}
	\caption{The ability of the proposed descriptors in the discrimination of texture classes. Left, we see texture images from two classes and their respective descriptors. Right, the descriptors are plotted in a same graph, showing visually the high discrimination potential.}
	\label{fig:discrim}
\end{figure}

By joining the power of fractal theory and more specifically fractal descriptors in the description of natural textures and the efficiency of the color approach described in Hoang et al. \cite{HGS05} we obtain a powerful descriptor for natural colored textures. Such descriptor is capable of capturing complex patterns in the texture which are capital for a complete and precise identification of a real scene. The Figure \ref{fig:discrim} shows in a simple example the potentiality of texture discrimination present in the proposed descriptor which may be seen even visually.

\section{Experiments}

\begin{figure}[htbp]
	\centering
		\includegraphics[width=\textwidth]{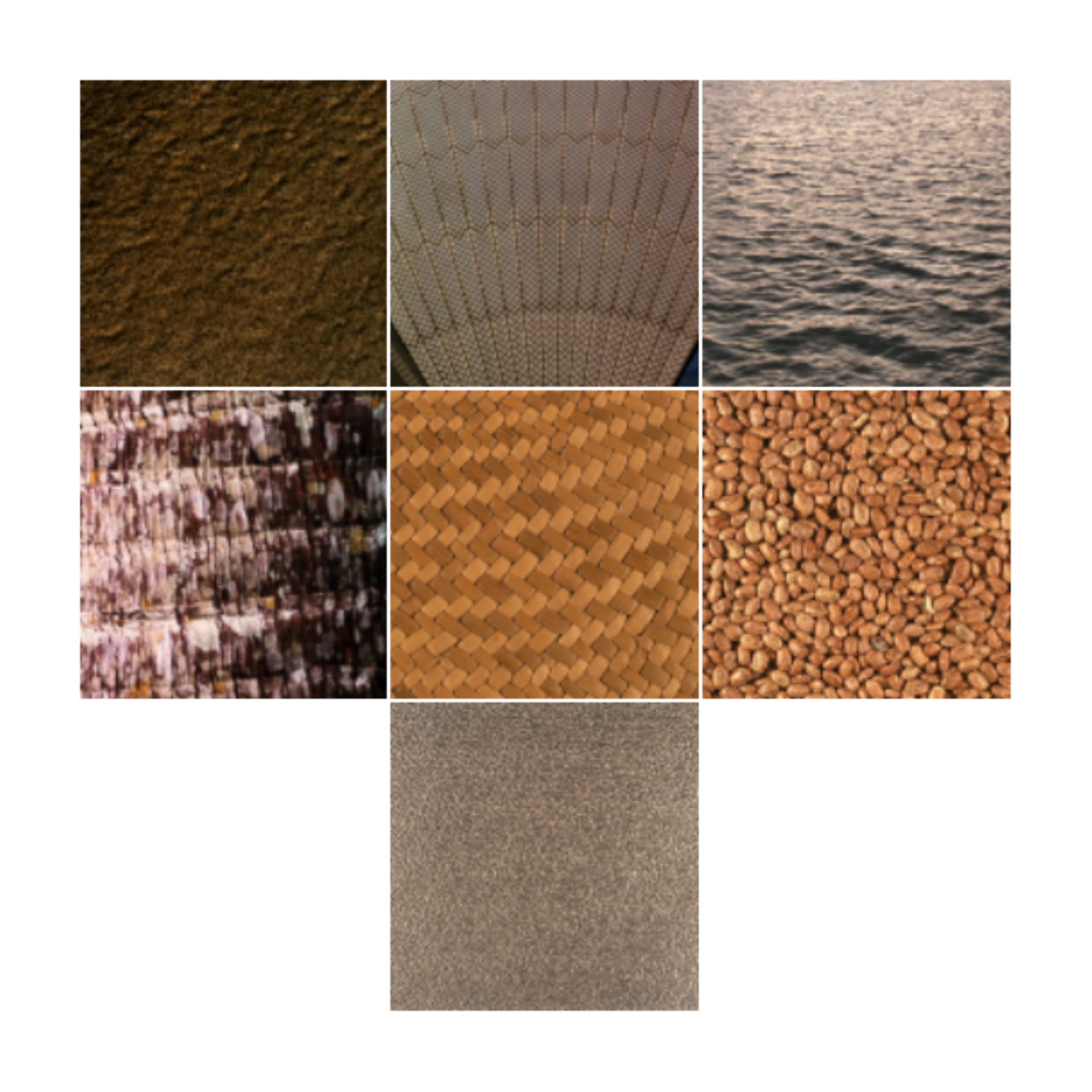}
	\caption{One image sample from each class of the Vistex dataset. From up to down, left to right: Sand, Tile, Water, Bark, Fabric, Food and Metal.}
	\label{fig:vistex}
\end{figure}

\begin{figure}[htbp]
	\centering
		\includegraphics[width=\textwidth]{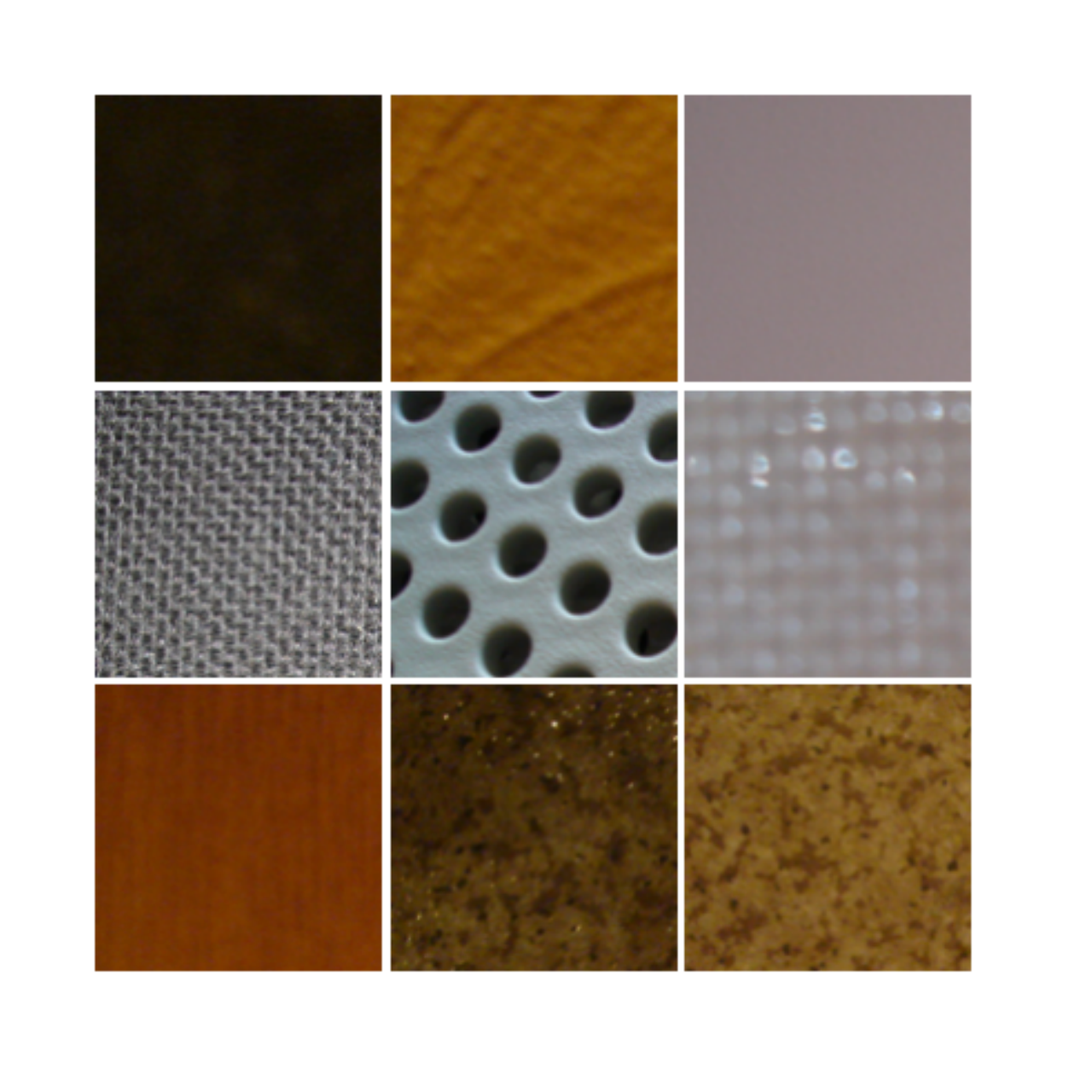}
	\caption{Some image samples (one from each class) from the USPTex dataset.}
	\label{fig:usptex}
\end{figure}

The performance of the proposed technique is tested by the classification of samples from VisTex \cite{Vistex}, a classical dataset of colored textures and USPTex, a dataset developed in the research group of the authors, which is composed by images of natural textures, photographed in high resolution. The Figures \ref{fig:vistex} and \ref{fig:usptex} show some image samples from each dataset.

The dimensionality of proposed descriptors is proportional to the dimension of the image. Thus, for the VisTex dataset, the signature are composed by 120 descriptors, while in USPTex we use 132 descriptors. We have not used other resolutions for descriptors, once such approach may compromise the fractality measure, in which all multiscale levels has equal importance for the description of the texture.

The results obtained were compared to the use of other classical methods for descriptors of colored textures, that is, color Gabor \cite{HGS05}, color histogram ratio \cite{PP03} and chromaticity moments \cite{P98}. The descriptors are classified by the well known K-Nearest Neighbor (KNN) method \cite{DH00}, with $k = 1$ (empirically determined) and using a 10-fold cross-validation process. The comparison is done in terms of the correctness rate with its confidence interval and the confusion matrices for each descriptor and dataset.

\section{Results}

In a first moment, we tested the classification performance of some metrics extracted from MFD curve, as it is done in the original MFD work \cite{MCSM02}. Here, we employed some fractal and statistical metrics, e.g., fractal dimension of channells 1, 2 and 3, mean, standard deviation, kurtosis, skewness, second and third order moments and the combination of these measures. The Table \ref{tab:stat} shows the correctness rate and associated error for each approach in Vistex dataset.
\begin{table*}[!htpb]
	\centering
	\caption{Percentage correctness rate and respective confidence interval in the classification of Vistex dataset using statistical metrics extracted from MFD curve.}	
		\begin{tabular}{ll}
			\hline\noalign{\smallskip}
                 Metric           & Correctness rate\\
                 \noalign{\smallskip}\hline\noalign{\smallskip}
								 FD Channell 1    & 30.51   $\pm$ 0.20\\                                  
								 FD Channell 2    & 33.11   $\pm$ 0.23\\                                  
								 FD Channell 3    & 35.38   $\pm$ 0.14\\                                  
                 Mean             & 32.46   $\pm$ 0.21\\
								 Std. Dev.        & 26.62   $\pm$ 0.17\\
                 Kurtosis         & 20.12   $\pm$ 0.27\\
                 Skewness         & 29.54   $\pm$ 0.39\\                 
                 2$^{th}$ Moment  & 26.62   $\pm$ 0.30\\                 
                 3$^{th}$ Moment  & 29.87   $\pm$ 0.40\\                                  
                 Combined         & 93.83   $\pm$ 0.21\\                 
                 Whole Curve      & \underline{95.12   $\pm$ 0.10}\\                 
								 \noalign{\smallskip}\hline
		\end{tabular}
	\label{tab:stat}
\end{table*} 
Thus, we see that the use of whole MFD curve in the composition of descriptors provided the best result. From now, we show results for the use of the whole curve in the tested datasets.

Initially, we see in the Table \ref{tab:result} the global correctness rate for each compared descriptor for Vistex and USPTex dataset. It is clearly noticeable that the proposed Fourier method presented the best result in the classification of both datasets. The proposed technique presented an advantage of 2.8\% in Vistex dataset and 2.7\% in USPTex over Gabor method, the second best technique in this experiment.
\begin{table*}[!htpb]
	\centering
	\caption{Percentage correctness rate and respective confidence interval in the classification of the tested datasets by the compared descriptors.}	
		\begin{tabular}{lll}
			\hline\noalign{\smallskip}
                 Method    & Vistex  & USPTex\\
                 \noalign{\smallskip}\hline\noalign{\smallskip}
                 Moment    & 68.83   $\pm$ 0.33 & 32.06   $\pm$ 0.05\\
								 Histogram & 78.89   $\pm$ 0.21 & 41.49   $\pm$ 0.18\\
                 Gabor     & 92.53   $\pm$ 0.18 & 86.47   $\pm$ 0.04\\
                 Fourier   & \underline{95.12   $\pm$ 0.10} & \underline{88.83   $\pm$ 0.07}\\                 
								 \noalign{\smallskip}\hline
		\end{tabular}
	\label{tab:result}
\end{table*} 

In the Table \ref{tab:matrices} we present the confusion matrices for each descriptor method. In this matrix, each raw (or column) represents a class and the value in raw $i$ and column $j$ expresses the number of objects of class $i$, but classified as being from class $j$. The ideal method (with a 100\% correctness rate) must present a diagonal confusion matrix. 

\begin{figure}[htbp]
	\centering
	\begin{tabular}{cc}
		\includegraphics[width=0.5\textwidth]{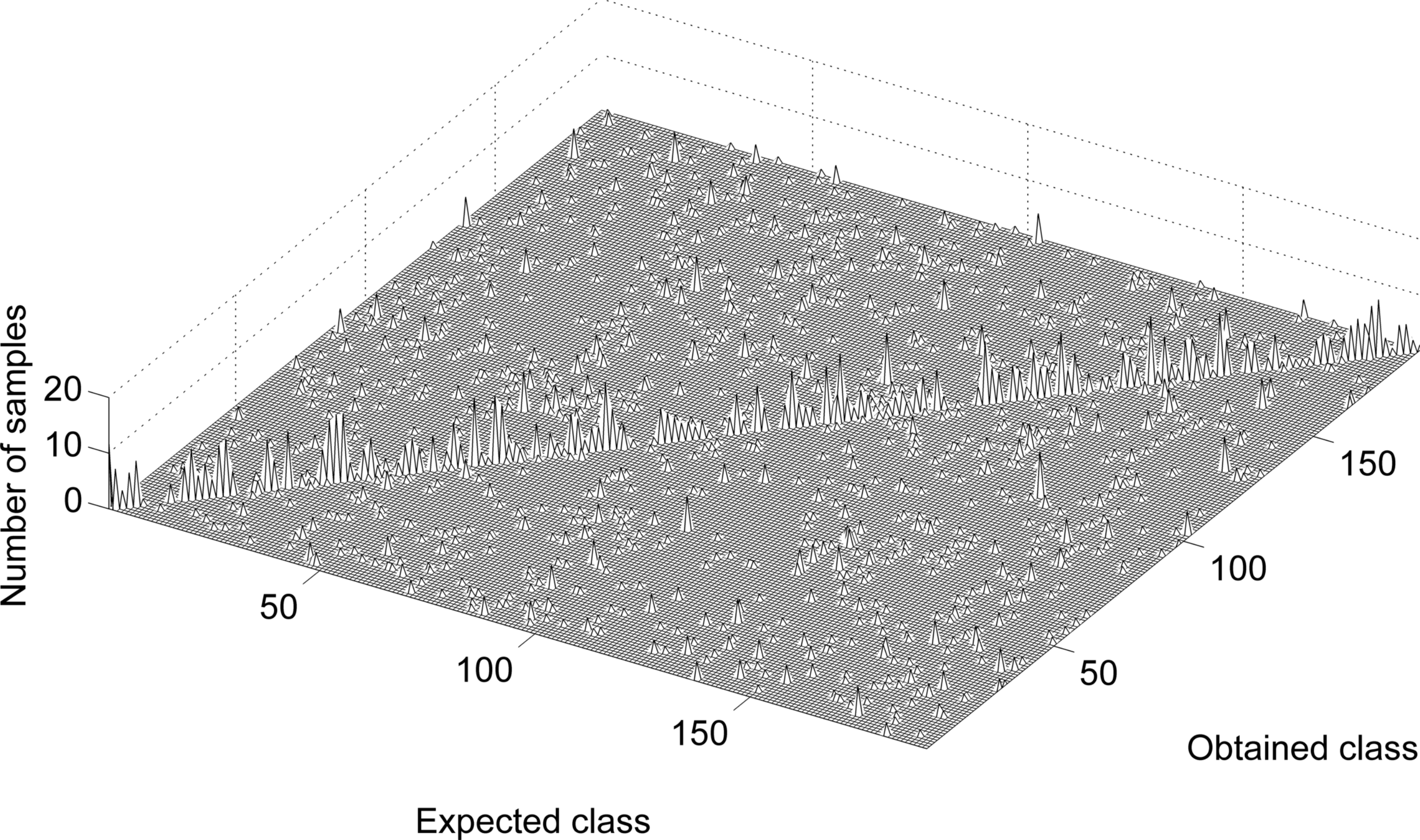} &       	\includegraphics[width=0.5\textwidth]{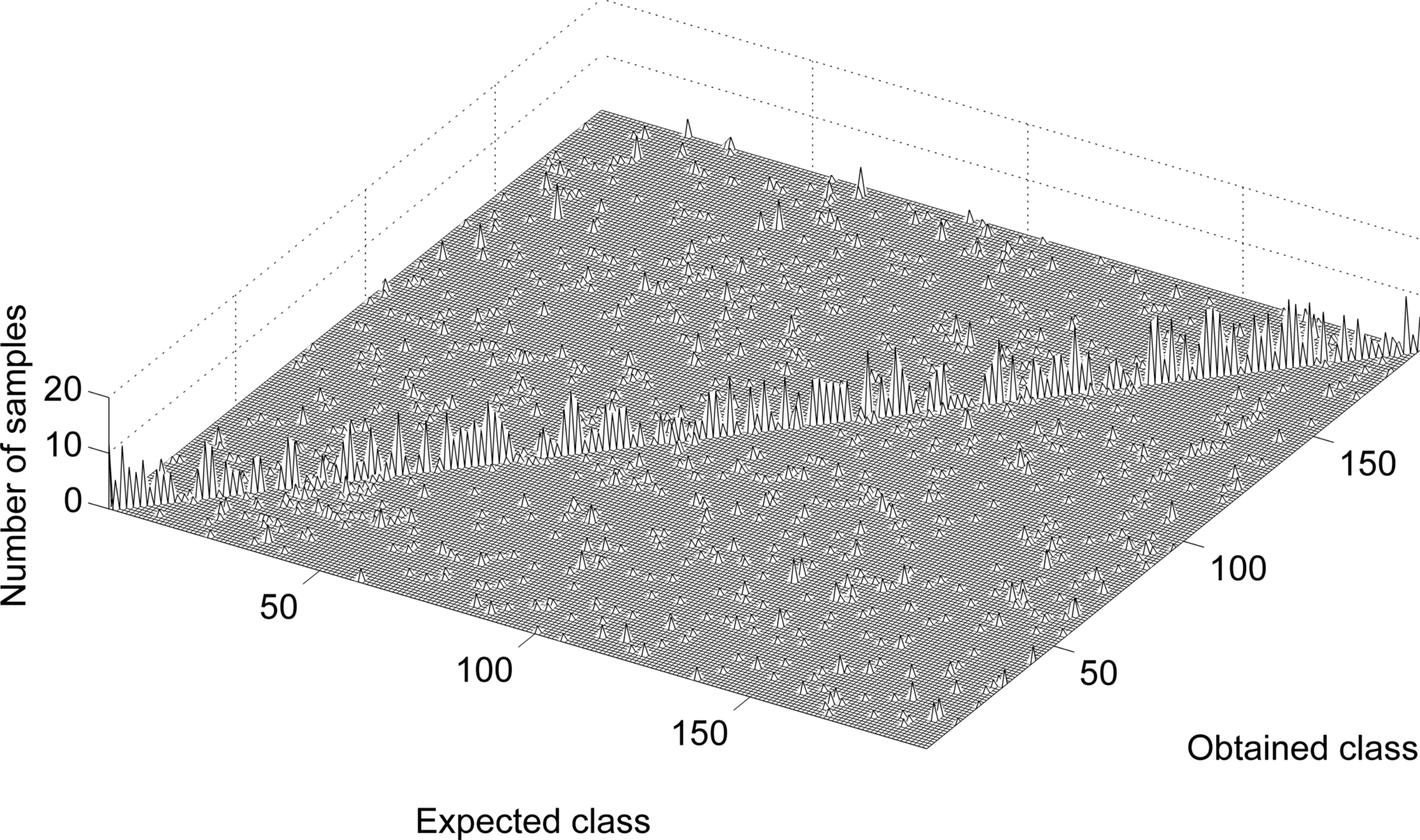}\\
		(a) & (b)\\
		 \includegraphics[width=0.5\textwidth]{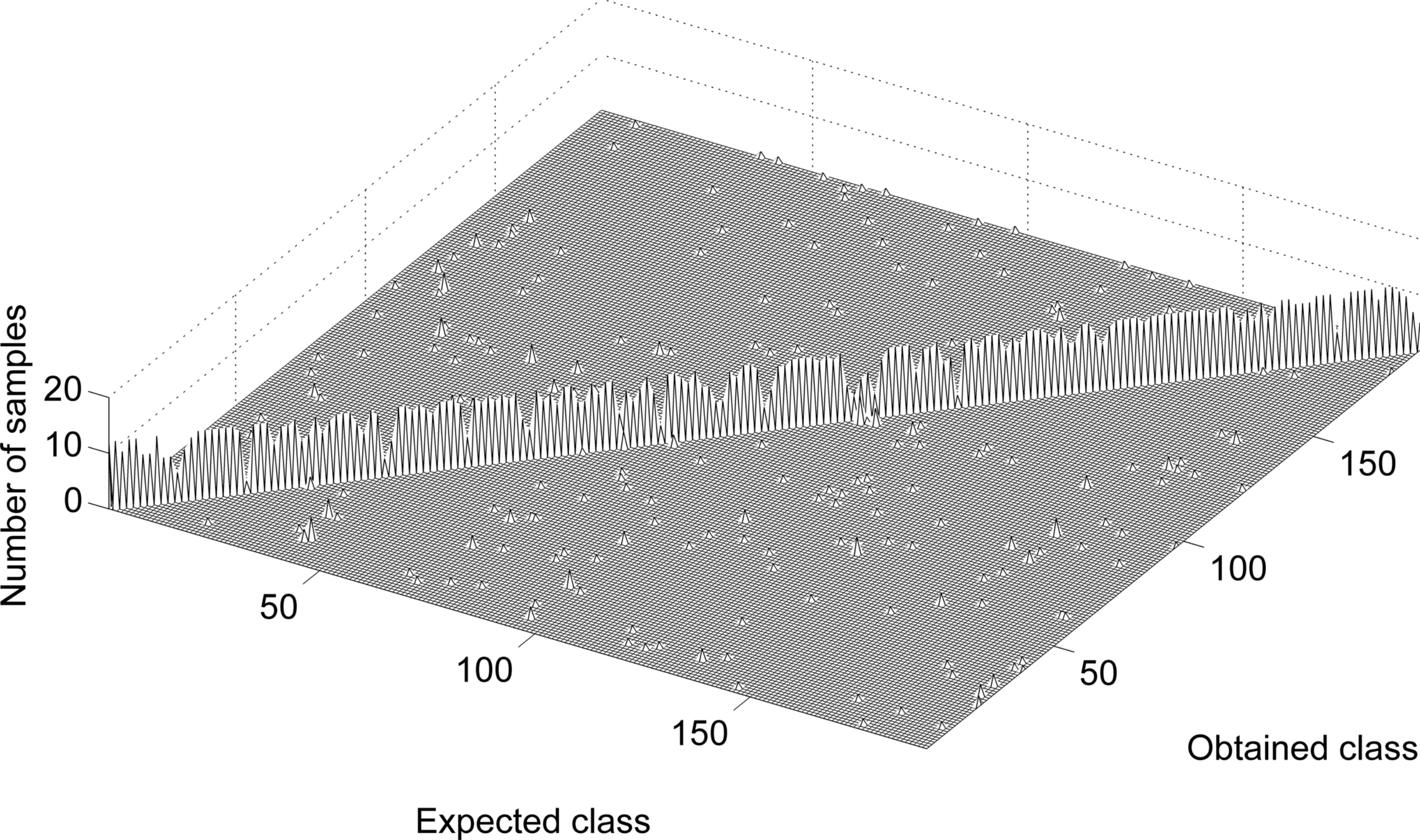} & \includegraphics[width=0.5\textwidth]{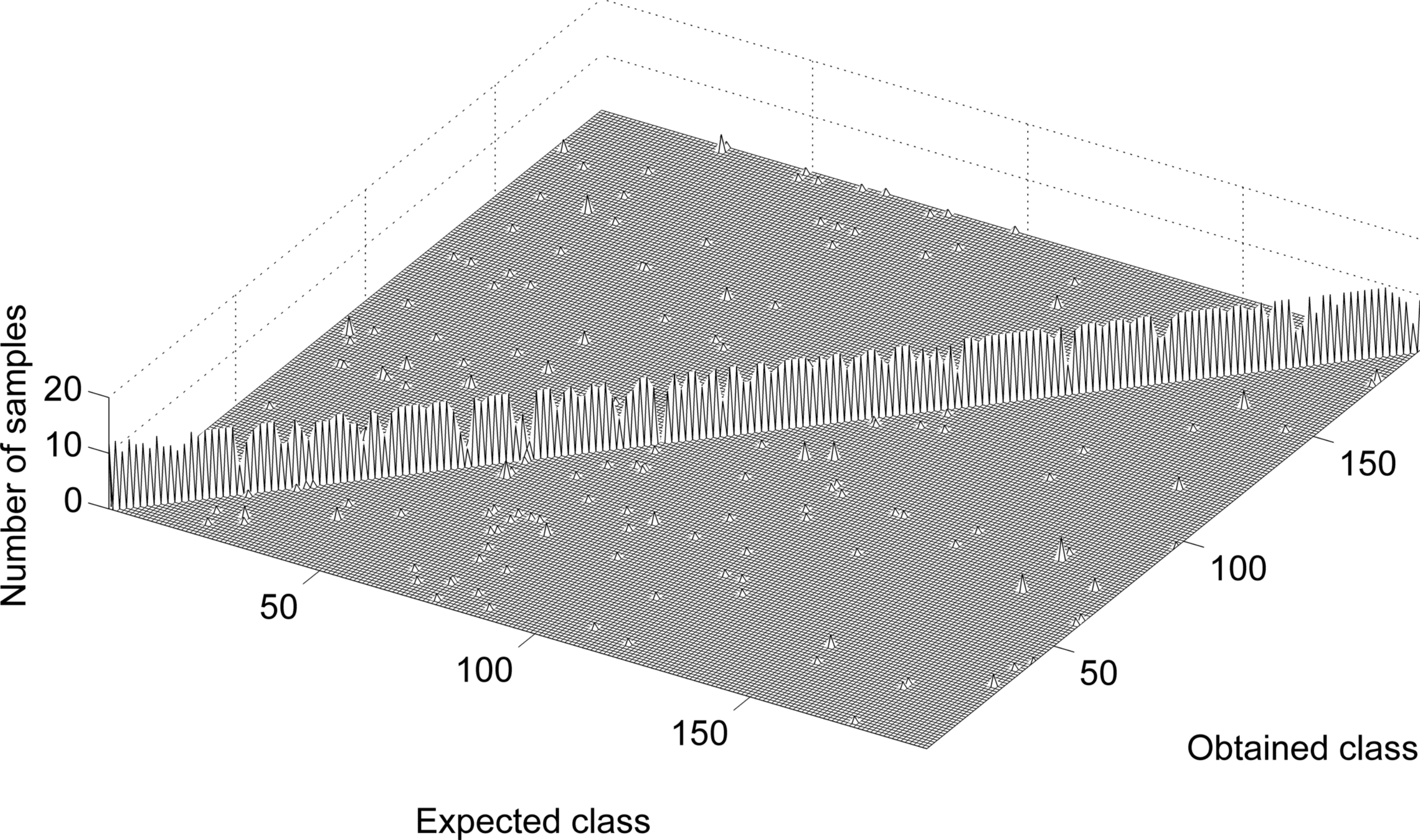}\\
		(c) & (d)\\
	\end{tabular}
	\caption{Surface visualization of confusion matrices for the classification of USPTex dataset using the compared descriptors. (a) Chromaticity moment. (b) Histogram. (c) Gabor. (d) Fourier.}
	\label{fig:matrices}
\end{figure}

For the Vistex dataset, the numbers correspond to the following classes (exemplified in the Figure \ref{fig:vistex}): 1-Bark, 2-Fabric, 3-Food, 4-Metal, 5-Sand, 6-Tile and 7-Water. As expected, the greater the correctness rate, more the confusion matrix presents diagonal aspect. Particularly, the proposed descriptor presented its best performance in the classes 3, 4 and 5. Good results are also observed in classes 1 and 6. Specially, in class 6, the proposed technique presented a relevant advantage over Gabor method. Gabor misclassified elements from class 6 as being from class 1, 2 and 5. This is explained by the self-similarity present in these classes. The fractal method, as expected, captured more faithfully the self-similar nuances.
\begin{table*}[!htpb]
	\centering
	
		\begin{tabular}{cc}
		\begin{tabular}{ccccccc}
                 \hline
25 & 5 & 2 & 3 & 4 & 13 & 0 \\
 10 & 61 & 3 & 3 & 1 & 2 & 0 \\
  1 & 1 & 43 & 0 & 1 & 2 & 0 \\
  1 & 5 & 0 & 16 & 0 & 0 & 2 \\
  1 & 2 & 2 & 0 & 20 & 3 & 0 \\
 10 & 3 & 5 & 0 & 4 & 22 & 0 \\
  2 & 1 & 0 & 4 & 0 & 0 & 25 \\                 
								 \hline			
		\end{tabular} &
				\begin{tabular}{ccccccc}
                 \hline
31 & 9 & 4 & 0 & 6 & 0 & 2 \\
  4 & 65 & 1 & 5 & 4 & 1 & 0 \\
  1 & 1 & 43 & 0 & 3 & 0 & 0 \\
  0 & 4 & 0 & 19 & 0 & 0 & 1 \\
  1 & 5 & 0 & 0 & 22 & 0 & 0 \\
  2 & 0 & 1 & 0 & 0 & 39 & 2 \\
  2 & 3 & 0 & 2 & 0 & 1 & 24 \\                
								 \hline			
		\end{tabular} \\
		(a) & (b) \\
		\begin{tabular}{ccccccc}
                 \hline
45 & 3 & 0 & 1 & 2 & 1 & 0 \\
  1 & 77 & 0 & 0 & 0 & 2 & 0 \\
  0 & 0 & 48 & 0 & 0 & 0 & 0 \\
  0 & 1 & 0 & 23 & 0 & 0 & 0 \\
  0 & 0 & 0 & 0 & 28 & 0 & 0 \\
  6 & 2 & 0 & 2 & 1 & 33 & 0 \\
  0 & 1 & 0 & 0 & 0 & 0 & 31 \\               
								 \hline			
		\end{tabular} &
		\begin{tabular}{ccccccc}
                 \hline
50 & 1 & 0 & 0 & 0 & 1 & 0 \\
  1 & 74 & 1 & 0 & 1 & 1 & 2 \\
  0 & 0 & 48 & 0 & 0 & 0 & 0 \\
  0 & 0 & 0 & 24 & 0 & 0 & 0 \\
  0 & 0 & 0 & 0 & 28 & 0 & 0 \\
  1 & 1 & 0 & 0 & 0 & 42 & 0 \\
  2 & 3 & 0 & 0 & 0 & 0 & 27 \\								 
                 \hline			
		\end{tabular}\\				
		(c) & (d)\\
	\end{tabular}	
	\caption{Confusion matrices for the classification of Vistex dataset using the compared descriptors. (a) Chromaticity moment. (b) Histogram. (c) Gabor. (d) Fourier.}
	\label{tab:matrices}
\end{table*} 
The confusion matrices from USPTex were represented in a different graphical manner, by using surface figures in the Figure \ref{fig:matrices}. In this representation, each position in the matrix is represented by a surface point and the height of the point, according to the legend on the axis, determines the value in that position. Looking at each figure, we observe that the matrix from the Fourier method presented a more continuous diagonal with the highest points, justifying the greater number of samples correctly classified. The Gabor method presented bad results around the class 110.

\section{Conclusion}

This work proposed a novel technique for the calculus of descriptors for colored textures. The method uses the Fourier spectral dimension associated to the spatial color transform proposed in Geusebroek et al. \cite{GRAA00}. Initially, the transform is applied to the original texture images in RGB space. Following, the classical Fourier transform is applied to the image and the values in the curve $log(power spectrum) \times log(frequency)$ are used as descriptors for the texture.

The accuracy of the proposed method was verified by a comparison with other classical techniques in color texture analysis. The results demonstrated the power of the novel technique once Fourier descriptors presented the greater precision in the classification of two complex datasets. Specially, the proposed technique showed great efficiency in capturing self-similar patterns in the texture.

The results suggest strongly that color Fourier fractal descriptors are an interesting alternative to be used in the solution of problems in which the description of an object is primordial, like tasks involving segmentation and classification of objects represented by their texture.

\section{Acknowledgements}
\label{sec:Acknowledgements}
Odemir M. Bruno gratefully acknowledges the financial support of CNPq (National Council for Scientific and Technological Development, Brazil) (Grant \#308449/2010-0 and \#473893/2010-0) and FAPESP (The State of S\~ao Paulo Research Foundation) (Grant \# 2011/01523-1). Jo\~ao B. Florindo is grateful to CNPq(National Council for Scientific and Technological Development, Brazil) for his doctorate grant.

\newpage


\begin{thebibliography}{29}%
\makeatletter
\providecommand \@ifxundefined [1]{%
 \@ifx{#1\undefined}
}%
\providecommand \@ifnum [1]{%
 \ifnum #1\expandafter \@firstoftwo
 \else \expandafter \@secondoftwo
 \fi
}%
\providecommand \@ifx [1]{%
 \ifx #1\expandafter \@firstoftwo
 \else \expandafter \@secondoftwo
 \fi
}%
\providecommand \natexlab [1]{#1}%
\providecommand \enquote  [1]{``#1''}%
\providecommand \bibnamefont  [1]{#1}%
\providecommand \bibfnamefont [1]{#1}%
\providecommand \citenamefont [1]{#1}%
\providecommand \href@noop [0]{\@secondoftwo}%
\providecommand \href [0]{\begingroup \@sanitize@url \@href}%
\providecommand \@href[1]{\@@startlink{#1}\@@href}%
\providecommand \@@href[1]{\endgroup#1\@@endlink}%
\providecommand \@sanitize@url [0]{\catcode `\\12\catcode `\$12\catcode
  `\&12\catcode `\#12\catcode `\^12\catcode `\_12\catcode `\%12\relax}%
\providecommand \@@startlink[1]{}%
\providecommand \@@endlink[0]{}%
\providecommand \url  [0]{\begingroup\@sanitize@url \@url }%
\providecommand \@url [1]{\endgroup\@href {#1}{\urlprefix }}%
\providecommand \urlprefix  [0]{URL }%
\providecommand \Eprint [0]{\href }%
\providecommand \doibase [0]{http://dx.doi.org/}%
\providecommand \selectlanguage [0]{\@gobble}%
\providecommand \bibinfo  [0]{\@secondoftwo}%
\providecommand \bibfield  [0]{\@secondoftwo}%
\providecommand \translation [1]{[#1]}%
\providecommand \BibitemOpen [0]{}%
\providecommand \bibitemStop [0]{}%
\providecommand \bibitemNoStop [0]{.\EOS\space}%
\providecommand \EOS [0]{\spacefactor3000\relax}%
\providecommand \BibitemShut  [1]{\csname bibitem#1\endcsname}%
\let\auto@bib@innerbib\@empty
\bibitem [{\citenamefont {Mandelbrot}(1975)}]{M75}%
  \BibitemOpen
  \bibfield  {author} {\bibinfo {author} {\bibfnamefont {B.~B.}\ \bibnamefont
  {Mandelbrot}},\ }\href@noop {} {\emph {\bibinfo {title} {The Fractal Geometry
  of Nature}}}\ (\bibinfo  {publisher} {Freeman},\ \bibinfo {address} {NY,
  USA},\ \bibinfo {year} {1975})\BibitemShut {NoStop}%
\bibitem [{\citenamefont {Manoel}\ \emph {et~al.}(2002)\citenamefont {Manoel},
  \citenamefont {da~Fontoura~Costa}, \citenamefont {Streicher},\ and\
  \citenamefont {M{\"u}ller}}]{MCSM02}%
  \BibitemOpen
  \bibfield  {author} {\bibinfo {author} {\bibfnamefont {E.~T.~M.}\
  \bibnamefont {Manoel}}, \bibinfo {author} {\bibfnamefont {L.}~\bibnamefont
  {da~Fontoura~Costa}}, \bibinfo {author} {\bibfnamefont {J.}~\bibnamefont
  {Streicher}}, \ and\ \bibinfo {author} {\bibfnamefont {G.~B.}\ \bibnamefont
  {M{\"u}ller}},\ }\bibfield  {title} {\enquote {\bibinfo {title} {Multiscale
  fractal characterization of three-dimensional gene expression data},}\ }in\
  \href@noop {} {\emph {\bibinfo {booktitle} {SIBGRAPI}}}\ (\bibinfo
  {publisher} {IEEE Computer Society},\ \bibinfo {year} {2002})\ pp.\ \bibinfo
  {pages} {269--274}\BibitemShut {NoStop}%
\bibitem [{\citenamefont {Bruno}\ \emph {et~al.}(2008)\citenamefont {Bruno},
  \citenamefont {de~Oliveira~Plotze}, \citenamefont {Falvo},\ and\
  \citenamefont {de~Castro}}]{BPFC08}%
  \BibitemOpen
  \bibfield  {author} {\bibinfo {author} {\bibfnamefont {O.~M.}\ \bibnamefont
  {Bruno}}, \bibinfo {author} {\bibfnamefont {R.}~\bibnamefont
  {de~Oliveira~Plotze}}, \bibinfo {author} {\bibfnamefont {M.}~\bibnamefont
  {Falvo}}, \ and\ \bibinfo {author} {\bibfnamefont {M.}~\bibnamefont
  {de~Castro}},\ }\bibfield  {title} {\enquote {\bibinfo {title} {Fractal
  dimension applied to plant identification},}\ }\href@noop {} {\bibfield
  {journal} {\bibinfo  {journal} {Information Sciences}\ }\textbf {\bibinfo
  {volume} {178}},\ \bibinfo {pages} {2722--2733} (\bibinfo {year}
  {2008})}\BibitemShut {NoStop}%
\bibitem [{\citenamefont {Materka}\ \emph {et~al.}(1998)\citenamefont
  {Materka}, \citenamefont {Strzelecki}, \citenamefont {Analysis},
  \citenamefont {Review}, \citenamefont {Materka},\ and\ \citenamefont
  {Strzelecki}}]{MS98}%
  \BibitemOpen
  \bibfield  {author} {\bibinfo {author} {\bibfnamefont {A.}~\bibnamefont
  {Materka}}, \bibinfo {author} {\bibfnamefont {M.}~\bibnamefont {Strzelecki}},
  \bibinfo {author} {\bibfnamefont {T.}~\bibnamefont {Analysis}}, \bibinfo
  {author} {\bibfnamefont {M.~A.}\ \bibnamefont {Review}}, \bibinfo {author}
  {\bibfnamefont {A.}~\bibnamefont {Materka}}, \ and\ \bibinfo {author}
  {\bibfnamefont {M.}~\bibnamefont {Strzelecki}},\ }\href@noop {} {\enquote
  {\bibinfo {title} {Texture analysis methods - a review},}\ }\bibinfo {type}
  {Tech. Rep.}\ (\bibinfo  {institution} {Institute of Electronics, Technical
  University of Lodz},\ \bibinfo {year} {1998})\BibitemShut {NoStop}%
\bibitem [{\citenamefont {Quevedo}\ \emph {et~al.}(2008)\citenamefont
  {Quevedo}, \citenamefont {Mendoza}, \citenamefont {Aguilera}, \citenamefont
  {Chanona},\ and\ \citenamefont {Gutierrez-Lopez}}]{QMACG08}%
  \BibitemOpen
  \bibfield  {author} {\bibinfo {author} {\bibfnamefont {R.}~\bibnamefont
  {Quevedo}}, \bibinfo {author} {\bibfnamefont {F.}~\bibnamefont {Mendoza}},
  \bibinfo {author} {\bibfnamefont {J.~M.}\ \bibnamefont {Aguilera}}, \bibinfo
  {author} {\bibfnamefont {J.}~\bibnamefont {Chanona}}, \ and\ \bibinfo
  {author} {\bibfnamefont {G.}~\bibnamefont {Gutierrez-Lopez}},\ }\bibfield
  {title} {\enquote {\bibinfo {title} {{Determination of Senescent Spotting in
  Banana (Musa cavendish) Using Fractal Texture Fourier Image}},}\ }\href@noop
  {} {\bibfield  {journal} {\bibinfo  {journal} {{Journal of Food
  Engineering}}\ }\textbf {\bibinfo {volume} {{84}}},\ \bibinfo {pages}
  {{509--515}} (\bibinfo {year} {{2008}})}\BibitemShut {NoStop}%
\bibitem [{\citenamefont {Tian-Gang}, \citenamefont {Wang},\ and\ \citenamefont
  {Zhao}(2007)}]{TWZ07}%
  \BibitemOpen
  \bibfield  {author} {\bibinfo {author} {\bibfnamefont {L.}~\bibnamefont
  {Tian-Gang}}, \bibinfo {author} {\bibfnamefont {S.}~\bibnamefont {Wang}}, \
  and\ \bibinfo {author} {\bibfnamefont {N.}~\bibnamefont {Zhao}},\ }\bibfield
  {title} {\enquote {\bibinfo {title} {{Fractal Research of Pathological Tissue
  Images}},}\ }\href@noop {} {\bibfield  {journal} {\bibinfo  {journal}
  {{Computerized Medical Imaging and Graphics}}\ }\textbf {\bibinfo {volume}
  {{31}}},\ \bibinfo {pages} {{665--671}} (\bibinfo {year}
  {{2007}})}\BibitemShut {NoStop}%
\bibitem [{\citenamefont {Millan}\ and\ \citenamefont
  {Gonzalez-Posada}(2005)}]{MG05}%
  \BibitemOpen
  \bibfield  {author} {\bibinfo {author} {\bibfnamefont {H.}~\bibnamefont
  {Millan}}\ and\ \bibinfo {author} {\bibfnamefont {M.}~\bibnamefont
  {Gonzalez-Posada}},\ }\bibfield  {title} {\enquote {\bibinfo {title}
  {{Modelling Soil Water Retention Scaling. Comparison of a Classical Fractal
  Model with a Piecewise Approach}},}\ }\href@noop {} {\bibfield  {journal}
  {\bibinfo  {journal} {{GEODERMA}}\ }\textbf {\bibinfo {volume} {{125}}},\
  \bibinfo {pages} {{25--38}} (\bibinfo {year} {{2005}})}\BibitemShut {NoStop}%
\bibitem [{\citenamefont {Backes}, \citenamefont {Casanova},\ and\
  \citenamefont {Bruno}(2009)}]{BCB09}%
  \BibitemOpen
  \bibfield  {author} {\bibinfo {author} {\bibfnamefont {A.~R.}\ \bibnamefont
  {Backes}}, \bibinfo {author} {\bibfnamefont {D.}~\bibnamefont {Casanova}}, \
  and\ \bibinfo {author} {\bibfnamefont {O.~M.}\ \bibnamefont {Bruno}},\
  }\bibfield  {title} {\enquote {\bibinfo {title} {Plant leaf identification
  based on volumetric fractal dimension},}\ }\href@noop {} {\bibfield
  {journal} {\bibinfo  {journal} {International Journal of Pattern Recognition
  and Artificial Intelligence (IJPRAI)}\ }\textbf {\bibinfo {volume} {23}},\
  \bibinfo {pages} {1145--1160} (\bibinfo {year} {2009})}\BibitemShut {NoStop}%
\bibitem [{\citenamefont {Plotze}\ \emph {et~al.}(2005)\citenamefont {Plotze},
  \citenamefont {Padua}, \citenamefont {Falvo}, \citenamefont {Vieira},
  \citenamefont {Oliveira},\ and\ \citenamefont {Bruno}}]{PPFVOB05}%
  \BibitemOpen
  \bibfield  {author} {\bibinfo {author} {\bibfnamefont {R.~O.}\ \bibnamefont
  {Plotze}}, \bibinfo {author} {\bibfnamefont {J.~G.}\ \bibnamefont {Padua}},
  \bibinfo {author} {\bibfnamefont {M.}~\bibnamefont {Falvo}}, \bibinfo
  {author} {\bibfnamefont {M.~L.~C.}\ \bibnamefont {Vieira}}, \bibinfo {author}
  {\bibfnamefont {G.~C.~X.}\ \bibnamefont {Oliveira}}, \ and\ \bibinfo {author}
  {\bibfnamefont {O.~M.}\ \bibnamefont {Bruno}},\ }\bibfield  {title} {\enquote
  {\bibinfo {title} {Leaf shape analysis by the multiscale minkowski fractal
  dimension, a new morphometric method: a study in passiflora l.
  (passifloraceae)},}\ }\href@noop {} {\bibfield  {journal} {\bibinfo
  {journal} {Canadian Journal of Botany-Revue Canadienne de Botanique}\
  }\textbf {\bibinfo {volume} {83}},\ \bibinfo {pages} {287--301} (\bibinfo
  {year} {2005})}\BibitemShut {NoStop}%
\bibitem [{\citenamefont {Florindo}, \citenamefont {De~Castro},\ and\
  \citenamefont {Bruno}(2010)}]{FCB10}%
  \BibitemOpen
  \bibfield  {author} {\bibinfo {author} {\bibfnamefont {J.~B.}\ \bibnamefont
  {Florindo}}, \bibinfo {author} {\bibfnamefont {M.}~\bibnamefont {De~Castro}},
  \ and\ \bibinfo {author} {\bibfnamefont {O.~M.}\ \bibnamefont {Bruno}},\
  }\bibfield  {title} {\enquote {\bibinfo {title} {{Enhancing Multiscale
  Fractal Descriptors Using Functional Data Analysis}},}\ }\href@noop {}
  {\bibfield  {journal} {\bibinfo  {journal} {{International Journal of
  Bifurcation and Chaos}}\ }\textbf {\bibinfo {volume} {{20}}},\ \bibinfo
  {pages} {{3443--3460}} (\bibinfo {year} {{2010}})}\BibitemShut {NoStop}%
\bibitem [{\citenamefont {Avnir}\ \emph {et~al.}(1998)\citenamefont {Avnir},
  \citenamefont {Biham}, \citenamefont {Lidar},\ and\ \citenamefont
  {Malcai}}]{ABLM98}%
  \BibitemOpen
  \bibfield  {author} {\bibinfo {author} {\bibfnamefont {D.}~\bibnamefont
  {Avnir}}, \bibinfo {author} {\bibfnamefont {O.}~\bibnamefont {Biham}},
  \bibinfo {author} {\bibfnamefont {D.}~\bibnamefont {Lidar}}, \ and\ \bibinfo
  {author} {\bibfnamefont {O.}~\bibnamefont {Malcai}},\ }\bibfield  {title}
  {\enquote {\bibinfo {title} {{Applied Mathematics: Is the Geometry of Nature
  Fractal?}}}\ }\href@noop {} {\bibfield  {journal} {\bibinfo  {journal}
  {Science}\ }\textbf {\bibinfo {volume} {279}},\ \bibinfo {pages} {39--40}
  (\bibinfo {year} {1998})}\BibitemShut {NoStop}%
\bibitem [{\citenamefont {Mandelbrot}(1998)}]{M98}%
  \BibitemOpen
  \bibfield  {author} {\bibinfo {author} {\bibfnamefont {B.~B.}\ \bibnamefont
  {Mandelbrot}},\ }\bibfield  {title} {\enquote {\bibinfo {title} {{Is nature
  fractal?}}}\ }\href@noop {} {\bibfield  {journal} {\bibinfo  {journal}
  {Science.}\ }\textbf {\bibinfo {volume} {279}},\ \bibinfo {pages} {783--785}
  (\bibinfo {year} {1998})}\BibitemShut {NoStop}%
\bibitem [{\citenamefont {Carlin}(2000)}]{C00}%
  \BibitemOpen
  \bibfield  {author} {\bibinfo {author} {\bibfnamefont {M.}~\bibnamefont
  {Carlin}},\ }\bibfield  {title} {\enquote {\bibinfo {title} {Measuring the
  complexity of non-fractal shapes by a fractal method},}\ }\href@noop {}
  {\bibfield  {journal} {\bibinfo  {journal} {Pattern Recognition Letters}\
  }\textbf {\bibinfo {volume} {21}},\ \bibinfo {pages} {1013--1017} (\bibinfo
  {year} {2000})}\BibitemShut {NoStop}%
\bibitem [{\citenamefont {Russ}(1994)}]{R94}%
  \BibitemOpen
  \bibfield  {author} {\bibinfo {author} {\bibfnamefont {J.~C.}\ \bibnamefont
  {Russ}},\ }\href@noop {} {\emph {\bibinfo {title} {Fractal Surfaces}}}\
  (\bibinfo  {publisher} {Plenum Press},\ \bibinfo {address} {New York},\
  \bibinfo {year} {1994})\BibitemShut {NoStop}%
\bibitem [{\citenamefont {Manjunath}\ \emph {et~al.}(2001)\citenamefont
  {Manjunath}, \citenamefont {Ohm}, \citenamefont {Vasudevan},\ and\
  \citenamefont {Yamada}}]{MOVY01}%
  \BibitemOpen
  \bibfield  {author} {\bibinfo {author} {\bibfnamefont {B.}~\bibnamefont
  {Manjunath}}, \bibinfo {author} {\bibfnamefont {J.}~\bibnamefont {Ohm}},
  \bibinfo {author} {\bibfnamefont {V.}~\bibnamefont {Vasudevan}}, \ and\
  \bibinfo {author} {\bibfnamefont {A.}~\bibnamefont {Yamada}},\ }\bibfield
  {title} {\enquote {\bibinfo {title} {{Color and Texture Descriptors}},}\
  }\href@noop {} {\bibfield  {journal} {\bibinfo  {journal} {{IEEE Transactions
  on Circuits and Systems for video Technology}}\ }\textbf {\bibinfo {volume}
  {{11}}},\ \bibinfo {pages} {{703--715}} (\bibinfo {year}
  {{2001}})}\BibitemShut {NoStop}%
\bibitem [{\citenamefont {Geusebroek}\ \emph {et~al.}(2000)\citenamefont
  {Geusebroek}, \citenamefont {van~den Boomgaard}, \citenamefont {Smeulders},\
  and\ \citenamefont {Dev}}]{GRAA00}%
  \BibitemOpen
  \bibfield  {author} {\bibinfo {author} {\bibfnamefont {J.-M.}\ \bibnamefont
  {Geusebroek}}, \bibinfo {author} {\bibfnamefont {R.}~\bibnamefont {van~den
  Boomgaard}}, \bibinfo {author} {\bibfnamefont {A.~W.~M.}\ \bibnamefont
  {Smeulders}}, \ and\ \bibinfo {author} {\bibfnamefont {A.}~\bibnamefont
  {Dev}},\ }\bibfield  {title} {\enquote {\bibinfo {title} {Color and scale:
  The spatial structure of color images},}\ }in\ \href@noop {} {\emph {\bibinfo
  {booktitle} {Sixth Europian Conference on Computer Vision (ECCV}}}\ (\bibinfo
   {publisher} {Springer},\ \bibinfo {year} {2000})\ pp.\ \bibinfo {pages}
  {331--341}\BibitemShut {NoStop}%
\bibitem [{\citenamefont {Muneeswaran}\ \emph {et~al.}(2005)\citenamefont
  {Muneeswaran}, \citenamefont {Ganesan}, \citenamefont {Arumugam},\ and\
  \citenamefont {Soundar}}]{MGASR05}%
  \BibitemOpen
  \bibfield  {author} {\bibinfo {author} {\bibfnamefont {K.}~\bibnamefont
  {Muneeswaran}}, \bibinfo {author} {\bibfnamefont {L.}~\bibnamefont
  {Ganesan}}, \bibinfo {author} {\bibfnamefont {S.}~\bibnamefont {Arumugam}}, \
  and\ \bibinfo {author} {\bibfnamefont {K.~R.}\ \bibnamefont {Soundar}},\
  }\bibfield  {title} {\enquote {\bibinfo {title} {Texture classification with
  combined rotation and scale invariant wavelet features},}\ }\href@noop {}
  {\bibfield  {journal} {\bibinfo  {journal} {Pattern Recogn.}\ }\textbf
  {\bibinfo {volume} {38}},\ \bibinfo {pages} {1495--1506} (\bibinfo {year}
  {2005})}\BibitemShut {NoStop}%
\bibitem [{\citenamefont {Harte}(2001)}]{H01}%
  \BibitemOpen
  \bibfield  {author} {\bibinfo {author} {\bibfnamefont {D.}~\bibnamefont
  {Harte}},\ }\href@noop {} {\emph {\bibinfo {title} {Multifractals: theory and
  applications}}}\ (\bibinfo  {publisher} {Chapman and Hall/CRC},\ \bibinfo
  {year} {2001})\BibitemShut {NoStop}%
\bibitem [{\citenamefont {Falconer}(1986)}]{F86}%
  \BibitemOpen
  \bibfield  {author} {\bibinfo {author} {\bibfnamefont {K.~J.}\ \bibnamefont
  {Falconer}},\ }\href@noop {} {\emph {\bibinfo {title} {The Geometry of
  Fractal Sets}}}\ (\bibinfo  {publisher} {Cambridge University Press},\
  \bibinfo {address} {New York, NY, USA},\ \bibinfo {year} {1986})\BibitemShut
  {NoStop}%
\bibitem [{\citenamefont {{O. Brigham}}(1974)}]{B74}%
  \BibitemOpen
  \bibfield  {author} {\bibinfo {author} {\bibnamefont {{O. Brigham}}},\
  }\href@noop {} {\emph {\bibinfo {title} {The Fast Fourier Transform}}}\
  (\bibinfo  {publisher} {Prentice-Hall},\ \bibinfo {address} {NY, USA},\
  \bibinfo {year} {1974})\BibitemShut {NoStop}%
\bibitem [{\citenamefont {Caelli}\ and\ \citenamefont {Reye}(1993)}]{CD93}%
  \BibitemOpen
  \bibfield  {author} {\bibinfo {author} {\bibfnamefont {T.}~\bibnamefont
  {Caelli}}\ and\ \bibinfo {author} {\bibfnamefont {D.}~\bibnamefont {Reye}},\
  }\bibfield  {title} {\enquote {\bibinfo {title} {On the classification of
  image regions by colour, texture and shape},}\ }\href@noop {} {\bibfield
  {journal} {\bibinfo  {journal} {Pattern Recognition}\ }\textbf {\bibinfo
  {volume} {26}},\ \bibinfo {pages} {461 -- 470} (\bibinfo {year}
  {1993})}\BibitemShut {NoStop}%
\bibitem [{\citenamefont {Paschos}(1998)}]{P98}%
  \BibitemOpen
  \bibfield  {author} {\bibinfo {author} {\bibfnamefont {G.}~\bibnamefont
  {Paschos}},\ }\bibfield  {title} {\enquote {\bibinfo {title} {Chromatic
  correlation features for texture recognition},}\ }\href@noop {} {\bibfield
  {journal} {\bibinfo  {journal} {Pattern Recognition Letters}\ }\textbf
  {\bibinfo {volume} {19}},\ \bibinfo {pages} {643 -- 650} (\bibinfo {year}
  {1998})}\BibitemShut {NoStop}%
\bibitem [{\citenamefont {Jain}\ and\ \citenamefont {Healey}(1998)}]{JH98}%
  \BibitemOpen
  \bibfield  {author} {\bibinfo {author} {\bibfnamefont {A.}~\bibnamefont
  {Jain}}\ and\ \bibinfo {author} {\bibfnamefont {G.}~\bibnamefont {Healey}},\
  }\bibfield  {title} {\enquote {\bibinfo {title} {{A Multiscale Representation
  Including Opponent Color Features for Texture Recognition}},}\ }\href@noop {}
  {\bibfield  {journal} {\bibinfo  {journal} {{IEEE Transactions on Image
  Processing}}\ }\textbf {\bibinfo {volume} {{7}}},\ \bibinfo {pages}
  {{124--128}} (\bibinfo {year} {{1998}})}\BibitemShut {NoStop}%
\bibitem [{\citenamefont {Hoang}, \citenamefont {Geusebroek},\ and\
  \citenamefont {Smeulders}(2005)}]{HGS05}%
  \BibitemOpen
  \bibfield  {author} {\bibinfo {author} {\bibfnamefont {M.~A.}\ \bibnamefont
  {Hoang}}, \bibinfo {author} {\bibfnamefont {J.-M.}\ \bibnamefont
  {Geusebroek}}, \ and\ \bibinfo {author} {\bibfnamefont {A.~W.}\ \bibnamefont
  {Smeulders}},\ }\bibfield  {title} {\enquote {\bibinfo {title} {Color texture
  measurement and segmentation},}\ }\href@noop {} {\bibfield  {journal}
  {\bibinfo  {journal} {Signal Processing}\ }\textbf {\bibinfo {volume} {85}},\
  \bibinfo {pages} {265 -- 275} (\bibinfo {year} {2005})}\BibitemShut {NoStop}%
\bibitem [{\citenamefont {Witkin}(2003)}]{W84}%
  \BibitemOpen
  \bibfield  {author} {\bibinfo {author} {\bibfnamefont {A.~P.}\ \bibnamefont
  {Witkin}},\ }\bibfield  {title} {\enquote {\bibinfo {title} {Scale space
  filtering: a new approach to multi-scale descriptions},}\ }in\ \href@noop {}
  {\emph {\bibinfo {booktitle} {Proceedings...}}},\ \bibinfo {organization}
  {ICASSP - IEEE International Conference on Acoustics, Speech, and Signal
  Processing}\ (\bibinfo  {publisher} {GRETSI},\ \bibinfo {address} {Saint
  Martin d'Hères, France},\ \bibinfo {year} {2003})\ pp.\ \bibinfo {pages}
  {79--95}\BibitemShut {NoStop}%
\bibitem [{\citenamefont {da~F.~Costa}\ and\ \citenamefont {{Cesar,
  Jr.}}(2000)}]{CC00}%
  \BibitemOpen
  \bibfield  {author} {\bibinfo {author} {\bibfnamefont {L.}~\bibnamefont
  {da~F.~Costa}}\ and\ \bibinfo {author} {\bibfnamefont {R.~M.}\ \bibnamefont
  {{Cesar, Jr.}}},\ }\href@noop {} {\emph {\bibinfo {title} {Shape Analysis and
  Classification: Theory and Practice}}}\ (\bibinfo  {publisher} {CRC Press},\
  \bibinfo {year} {2000})\BibitemShut {NoStop}%
\bibitem [{\citenamefont {MIT}(2010)}]{Vistex}%
  \BibitemOpen
  \bibfield  {author} {\bibinfo {author} {\bibnamefont {MIT}},\ }\href
  {\linebreak
  http://vismod.media.mit.edu/vismod/imagery/VisionTexture/vistex.html}
  {\enquote {\bibinfo {title} {Mit vistex texture database},}\ } (\bibinfo
  {year} {2010})\BibitemShut {NoStop}%
\bibitem [{\citenamefont {Paschos}\ and\ \citenamefont {Petrou}(2003)}]{PP03}%
  \BibitemOpen
  \bibfield  {author} {\bibinfo {author} {\bibfnamefont {G.}~\bibnamefont
  {Paschos}}\ and\ \bibinfo {author} {\bibfnamefont {M.}~\bibnamefont
  {Petrou}},\ }\bibfield  {title} {\enquote {\bibinfo {title} {Histogram ratio
  features for color texture classification},}\ }\href@noop {} {\bibfield
  {journal} {\bibinfo  {journal} {Pattern Recognition Letters}\ }\textbf
  {\bibinfo {volume} {24}},\ \bibinfo {pages} {309 -- 314} (\bibinfo {year}
  {2003})}\BibitemShut {NoStop}%
\bibitem [{\citenamefont {Duda}\ and\ \citenamefont {Hart}(1973)}]{DH00}%
  \BibitemOpen
  \bibfield  {author} {\bibinfo {author} {\bibfnamefont {R.~O.}\ \bibnamefont
  {Duda}}\ and\ \bibinfo {author} {\bibfnamefont {P.~E.}\ \bibnamefont
  {Hart}},\ }\href@noop {} {\emph {\bibinfo {title} {Pattern Classification and
  Scene Analysis}}}\ (\bibinfo  {publisher} {Wiley},\ \bibinfo {address} {New
  York},\ \bibinfo {year} {1973})\BibitemShut {NoStop}%
\end{thebibliography}



%



\end{document}